\documentclass[aip,jcp,reprint,notitlepage,amsmath,amsfonts,amssymb,floatfix]{revtex4-1}
\usepackage{graphicx}
\usepackage{bm}
\usepackage{color}

\begin{document}
\title{Projected Coupled Cluster Theory}
\author{Yiheng Qiu}
\affiliation{Department of Chemistry, Rice University, Houston, TX 77005-1892}

\author{Thomas M. Henderson}
\affiliation{Department of Chemistry, Rice University, Houston, TX 77005-1892}
\affiliation{Department of Physics and Astronomy, Rice University, Houston, TX 77005-1892}

\author{Jinmo Zhao}
\affiliation{Department of Chemistry, Rice University, Houston, TX 77005-1892}

\author{Gustavo E. Scuseria}
\affiliation{Department of Chemistry, Rice University, Houston, TX 77005-1892}
\affiliation{Department of Physics and Astronomy, Rice University, Houston, TX 77005-1892}
\date{\today}

\begin{abstract}
Coupled cluster theory is the method of choice for weakly correlated systems. But in the strongly correlated regime, it faces a symmetry dilemma, where it either completely fails to describe the system, or has to artificially break certain symmetries. On the other hand, projected Hartree-Fock theory captures the essential physics of many kinds of strong correlations via symmetry breaking and restoration. In this work, we combine and try to retain the merits of these two methods by applying symmetry projection to broken symmetry coupled cluster wavefunctions. The non-orthogonal nature of states resulting from the application of symmetry projection operators furnishes particle-hole excitations to all orders, thus creating an obstacle for the exact evaluation of overlaps. Here we provide a solution via a disentanglement framework theory that can be approximated rigorously and systematically. Results of projected coupled cluster theory are presented for molecules and the Hubbard model, showing that spin projection significantly improves unrestricted coupled cluster theory while restoring good quantum numbers. The energy of projected coupled cluster theory reduces to the unprojected one in the thermodynamic limit, albeit at a much slower rate than projected Hartree-Fock.
\end{abstract}

\maketitle

\section{Introduction}
Single-reference coupled cluster theory (CC)\cite{Coester1958,Cizek1966,Paldus1999,Bartlett2007,ShavittBartlett} is often considered the gold standard of quantum chemistry.  While relatively expensive, coupled cluster with single and double excitations (CCSD) is affordable on systems of moderate to large size, and adding a perturbative correction for triple excitations leads to results which, for weakly correlated system, are generally within chemical accuracy at polynomial cost -- $\mathcal{O}(N^6)$ for CCSD and $\mathcal{O}(N^7)$ with perturbative triple excitation corrections, where $N$ is some measure of system size.

Unfortunately, single-reference coupled cluster's reputation is tarnished somewhat by its inability to describe strongly correlated problems with more than a few strongly correlated electrons.  Active space coupled cluster methods are some help here, but are not a panacea, and in general systems with many strongly correlated electrons cannot readily be described by symmetry-adapted coupled cluster techniques.  

To remedy this problem, we are often forced to use a broken-symmetry mean-field reference such as unrestricted Hartree-Fock (UHF), so that both the reference determinant and indeed the coupled cluster wave function lack some or perhaps all of the symmetries of the exact wave function.  In the thermodynamic limit, this symmetry breaking is in fact real and physical, but for finite systems it is artificial and should be avoided.  The practical result is that while symmetry-broken coupled cluster may be energetically accurate, the loss of symmetry can yield poor results for properties other than total energies. 

At the mean-field level, one can obtain the energetic benefits of symmetry breaking in a symmetry-adapted picture by using the projected Hartree-Fock (PHF) method.\cite{Lowdin55c,Ring80,Blaizot85,Schmid2004,PHF}  The idea of PHF is simple enough.   One allows symmetry breaking in the mean-field, and then projects the broken-symmetry determinant back onto the correct symmetry.  This can be accomplished by writing the PHF wave function as a relatively short linear combination of degenerate and non-orthogonal broken symmetry determinants obtainable from one another by a symmetry rotation operator.  PHF is a black-box technique without the necessity of picking active orbitals.  Furthermore, it possesses an underlying reference determinant -- in fact, a whole manifold of them -- which in principle allows for combination with single-reference coupled cluster theory.

Ideally, we would like to combine PHF and coupled cluster, and we have made several attempts to do so.\cite{Qiu2016,Qiu2017,Degroote2016,WahlenStrothman2016,Gomez2017,Hermes2017}  In our previous work, we have written the projected Hartree-Fock wave function in terms of particle-hole excitations acting on a symmetry-adapted reference, in which form we can readily combine PHF with fairly traditional symmetry-adapted coupled cluster theory.  In this work, we take a complementary approach and work in the symmetry-broken picture.  The idea is simply to carry out the symmetry projection of a broken-symmetry coupled cluster wave function.  We will focus on spin symmetry in this manuscript, because it is the symmetry that spontaneously breaks in molecular systems, but the basic framework is more general.  The main challenge for our projected coupled cluster theory is the non-orthogonality to which we have previously alluded.  Orthogonal determinants, differing by particle-hole excitations, lose their orthogonality after the symmetry rotation, but many of the simplifications of traditional coupled cluster theory rely on this orthogonality.  Practical approximations require some way of either evaluating or truncating the overlaps between nonorthogonal states.  In this manuscript, we do so by what we call disentangled cluster operators.  This formalism permits us to work with orthogonal particle-hole excitations and thereby truncate projected coupled cluster in a systematic way due to the decay of the disentangled cluster operators with excitation rank.

The projected coupled cluster theory (PCC) introduced in this manuscript has some nice features. If the cluster operators are set to zero, it reduces to PHF. If the projection operator is disregarded, it reduces to broken symmetry coupled cluster theory.  
In between, we have a theory with particle-hole excitations to all orders; this is the result of disentangling the action of symmetry projection and broken-symmetry coupled cluster, but is a model that can be rigorously approximated by truncation, very much in the spirit of traditional coupled cluster theory.
Just as the size-extensive component of the PHF energy is that of the symmetry-broken mean-field, the size-extensive component of the PCC energy is that of the symmetry-broken coupled cluster.  However, PCC reduces to broken symmetry coupled cluster at a much slower rate than PHF reduces to broken symmetry Hartree-Fock.  Moreover, it should not be forgotten that the PCC wave function has correct symmetry.  While the size extensive energetic component of PHF is the same as broken-symmetry mean field (though there is also a size-intensive component independent of system size in the thermodynamic limit\cite{PHF}), the wave functions are different.

Previously, Duguet proposed a symmetry broken and restored coupled cluster theory.\cite{Duguet2015} We have borrowed some concepts and nomenclature from his paper, but these two approaches are very different, as we shall examine below in more detail.

\section{Background}
Before we can discuss the basic formulation of our symmetry-projected coupled cluster theory, we briefly review both traditional coupled cluster theory and symmetry projection to establish some basic concepts.  Readers familiar with these ideas may freely skip the following two subsections, though we do use them to establish our notation.

\subsection{Traditional Coupled Cluster Theory}
In coupled cluster theory, the ground state wave function $|\Psi\rangle$ is approximated by the exponential ansatz 
\begin{equation}
|\Psi\rangle = \mathrm{e}^U |\phi\rangle
\end{equation}
where $|\phi\rangle$ is a single determinant and where $U$ is an excitation operator.  Inserting this wave function ansatz into the Schr\"odinger equation yields
\begin{equation}
H \, \mathrm{e}^U |\phi\rangle = E \, \mathrm{e}^U |\phi\rangle
\label{Eqn:Schrodinger}
\end{equation}
and leads to energy and amplitude equations
\begin{subequations}
\begin{align}
E &= \langle \phi| H \, \mathrm{e}^U |\phi\rangle,
\\
0 &= \langle \mu| \left(H - E\right) \, \mathrm{e}^U |\phi\rangle,
\end{align}
\label{Eqns:CC1}
\end{subequations}
where $|\mu\rangle$ is an excited determinant.  So-called unlinked terms appear in the foregoing equations but cancel out.\cite{Hirata}

Coupled cluster theory is not conventionally described as we have done.  More commonly, one multiplies both sides of Eqn. \ref{Eqn:Schrodinger} by $\exp(-U)$ to get a similarity-transformed Schr\"odinger equation:
\begin{equation}
\mathrm{e}^{-U} \, H \, \mathrm{e}^U | \phi \rangle = \bar{H} |\phi = E \, |\phi\rangle
\end{equation}
where generically the operator $\bar{\mathcal{O}}$ means
\begin{equation}
\bar{\mathcal{O}} = \mathrm{e}^{-U} \, \mathcal{O} \, \mathrm{e}^U.
\label{Eqn:DefBar}
\end{equation}
Now the energy and amplitudes defining $U$ are obtained from
\begin{subequations}
\begin{align}
E &= \langle \phi| \bar{H} |\phi\rangle,
\\
0 &= \langle \mu| \bar{H} |\phi\rangle.
\end{align}
\label{Eqns:CC2}
\end{subequations}
The energy and amplitude equations of Eqn. \ref{Eqns:CC2} together imply that $|\phi\rangle$ is a right-hand eigenstate of $\bar{H}$.  Since $\bar{H}$ is non-Hermitian, its left-hand eigenstate $\langle L|$ is not the adjoint of its right-hand eigenstate, and we can parametrize it in a configuration interaction-like way, as
\begin{equation}
\langle L| = \langle \phi| \left(1 + Z\right)
\end{equation}
where $Z$ is an excitation operator acting to the left, and thus a de-excitation operator acting to the right.  As the left-hand eigenstate, we can solve for $Z$ by demanding that
\begin{equation}
\langle \phi| \left(1 + Z\right) \bar{H} = E \, \langle \phi| \left(1 + Z\right).
\label{Eqn:LinResp}
\end{equation}
With a little effort, one can show that Eqns. \ref{Eqns:CC2} and \ref{Eqn:LinResp} together can be summarized as
\begin{subequations}
\begin{align}
\mathcal{E}
  &= \langle \phi| \left(1+Z\right) \, \bar{H} |\phi\rangle
   = \frac{\langle \phi| \left(1+Z\right) \, \bar{H} |\phi\rangle}{\langle \phi| 1 + Z|\phi\rangle},
\label{Eqn:CCEnergy}
\\
0 &= \frac{\partial \mathcal{E}}{\partial Z_\mu},
\label{Eqns:Tamplitudes}
\\
0 &= \frac{\partial \mathcal{E}}{\partial U_\mu},
\end{align}
\label{Eqns:CC3}
\end{subequations}
where $Z_\mu$ and $U_\mu$ are individual amplitudes in the operators $Z$ and $U$.  If the amplitudes defining $U$ satisfy Eqn. \ref{Eqns:Tamplitudes}, then for any amplitudes $Z$, the two energies $E$ and $\mathcal{E}$ are identical; otherwise, they may differ significantly.  We would argue that $\mathcal{E}$ is the more correct definition of the coupled cluster energy in general.

We have already noted that Eqns. \ref{Eqns:CC2} and $\ref{Eqns:CC3}$ yield equivalent formulations of coupled cluster theory.  Both are equivalent to the non-similarity-transformed approach of Eqns. \ref{Eqns:CC1}, as can be shown by making use of the facts that
\begin{subequations}
\begin{align}
\langle \phi| \mathrm{e}^{-U} &= \langle \phi|,
\\
\langle \mu| \mathrm{e}^{-U} &= \sum c_{\mu\nu} \, \langle \nu|
\end{align}
\label{Eqn:SimTrans2NonSimTrans}
\end{subequations}
where the coefficients $c_{\mu\nu}$ depend on the amplitudes defining $U$ and where the sum on $\nu$ includes determinants of equal or lower excitation level to $\langle \mu|$, including the ground state determinant $\langle\phi|$.

Thus far, we have made no approximations; when $U$ contains all possible excitations, the theory as we have outlined it is exact.  In practical calculations, $U$ and $Z$ must be truncated, and the resultant theories are named according to the excitation levels retained.  For example, the simplest theory is coupled cluster doubles (CCD) and retains only the double-excitation parts of $U$ and $Z$, where CCSD keeps both the single- and double-excitation parts of $U$ and $Z$.  Typically we truncate $Z$ at the same level to which we truncate $U$.

\subsection{Symmetry Projection
\label{Sec:SymmetryProjection}}
The basic idea of symmetry projection is straightforward.  From a broken symmetry wave function $|\psi\rangle$ the symmetry adapted component is simply
\begin{equation}
|\Psi\rangle = P |\psi\rangle
\end{equation}
where the projector $P$ is Hermitian and idempotent, and commutes with the Hamiltonian.

Some important symmetries, such as number and spin, are continuous.  For these symmetries, perhaps the simplest way to write the projector is as an integral over symmetry-generated transformations:
\begin{equation}
P = \frac{1}{V_\Omega} \, \int \mathrm{d}\Omega \, R(\Omega) \, w(\Omega)
\label{Eqn:Projector}
\end{equation}
where $R(\Omega)$ is a one-body rotation operator, $w(\Omega)$ is a weight which depends on the eigenvalue being projected onto, and $V_\Omega = \int \mathrm{d}\Omega$ is the volume of the space being integrated over.  If the symmetry is discrete, the integration is replaced by a summation.  In practice, the integral is replaced by a weighted sum over a grid even for a continuous symmetry such as spin.

Given a projected wave function, one has several options for extracting the energy.  In PHF we use a simple expectation value, but here we generalize to a biorthogonal expectation value as is used in coupled cluster theory (\textit{c.f.} Eqn. \ref{Eqn:CCEnergy}) because we will need this form for the projected coupled cluster energy.  In this kind of biorthogonal approach, one might write
\begin{equation}
E = \frac{\langle \chi|P^\dagger \, H \, P|\psi\rangle}{\langle \chi|P^\dagger \, P|\psi\rangle}
  = \frac{\langle \chi| P \, H|\psi\rangle}{\langle \chi | P |\psi\rangle},
\end{equation}
where $\langle \chi|$ is some other broken symmetry wave function and where we have used the properties of the projection operator.  In view of the integral form of the projection operator, the energy can be expressed in terms of reduced norm and Hamiltonian kernels $\mathcal{N}(\Omega)$ and $\mathcal{H}(\Omega)$:
\begin{subequations}
\begin{align}
\mathcal{N}(\Omega) &= \langle \chi| R(\Omega) |\psi\rangle,
\\
\mathcal{H}(\Omega) &= \frac{\langle \chi |R(\Omega) \, H |\psi\rangle}{\mathcal{N}(\Omega)},
\label{Def:HKernel}
\\
E &= \frac{\int \mathrm{d}\Omega \, w(\Omega) \, \mathcal{H}(\Omega) \, \mathcal{N}(\Omega)}{\int \mathrm{d}\Omega \, w(\Omega) \, \mathcal{N}(\Omega)}.
\end{align}
\label{Eqn:GeneralProjectedEnergy}
\end{subequations}
Note that $\langle \chi|R(\Omega) \, H |\psi\rangle$ is proportional to the norm kernel.

Our primary interest in this work is in spin projection.  While the general case of spin projection is slightly more complicated than what we have outlined above,\cite{PHF} for the special case of spin projection onto a singlet state we can write a projection-like operator
\begin{equation}
P_S = \int_0^{2\pi} \frac{\mathrm{d}\alpha}{2\pi} \, \int_0^{\pi} \frac{\sin(\beta) \, \mathrm{d}\beta}{2} \, \int_0^{2\pi} \frac{\mathrm{d}\gamma}{2\pi} R(\alpha,\beta,\gamma)
\end{equation}
where the rotation operator is
\begin{equation}
R(\alpha,\beta,\gamma) = \mathrm{e}^{-\mathrm{i} \, \alpha \, S_z} \, \mathrm{e}^{-\mathrm{i} \, \beta \, S_y} \, \mathrm{e}^{-\mathrm{i} \, \gamma \, S_z}.
\end{equation}
This projector simplifies to
\begin{equation}
P_S = P_{s_z = 0} \, \int_0^{\pi} \frac{\sin(\beta) \, \mathrm{d}\beta}{2} \, \mathrm{e}^{-\mathrm{i} \, \beta \, S_y} \, P_{s_z = 0}
\end{equation}
where $P_{s_z = 0}$ is the projector onto eigenstates of $S_z$ with eigenvalue zero.

When working in the unrestricted framework in which wave functions are eigenfunctions of $S_z$ but not of $S^2$, the projector simplifies somewhat.  In particular, we can use the facts that
\begin{subequations}
\begin{align}
P_{s_z = 0} |\psi\rangle &= |\psi\rangle,
\\
\langle \chi| P_{s_z = 0} &= \langle \chi|
\end{align}
\end{subequations}
for spin unrestricted wave functions $|\psi\rangle$ and $\langle \chi|$ to write the norm kernel, Hamiltonian kernel, and projected energy of Eqn. \ref{Eqn:GeneralProjectedEnergy} as simply
\begin{subequations}
\begin{align}
\mathcal{N}(\beta) &= \langle \chi | \mathrm{e}^{-\mathrm{i} \, \beta \, S_y} |\psi\rangle,
\\
\mathcal{H}(\beta) &= \frac{\langle \chi | \mathrm{e}^{-\mathrm{i} \, \beta \, S_y} \, H |\psi\rangle}{\langle \chi | \mathrm{e}^{-\mathrm{i} \, \beta \, S_y} |\psi\rangle},
\\
E &= \frac{\int_0^\pi \mathrm{d}\beta \, \sin(\beta) \, \mathcal{H}(\beta) \, \mathcal{N}(\beta)}{\int_0^\pi \mathrm{d}\beta \, \sin(\beta) \, \mathcal{N}(\beta)}.
\end{align}
\end{subequations}

\section{Projected Coupled Cluster Theory}
Now that we have reviewed the background material, we are in a position to introduce our spin-projected unrestricted coupled cluster theory.  In this section we make no approximations other than the truncation of the cluster operator.  As we shall see, the exact theory is computationally too cumbersome for practical use, and we will introduce computationally tractable approximate versions in Sec. \ref{Sec:Approximations}.  These approximations are motivated by the smallness of the cluster amplitudes in broken symmetry coupled cluster and afford truncation in the same spirit as traditional single-reference coupled cluster theory.

\subsection{Theory}
The basic idea of our projected coupled cluster (PCC) theory is simple: the symmetry adapted wave function $|\Psi\rangle$ is obtained by projecting a broken-symmetry coupled cluster wave function as
\begin{equation}
|\Psi\rangle = P \, \mathrm{e}^{U} |\phi\rangle.
\end{equation}

Just as for traditional coupled cluster, the PCC energy and amplitude equations are derived by inserting the wave function ansatz into the Schr\"odinger equation to get
\begin{subequations}
\begin{align}
E &= \frac{\langle \phi| P \, H \, \mathrm{e}^{U} |\phi\rangle}{\langle \phi| P \, \mathrm{e}^{U} |\phi\rangle},
\label{Eqn:ProjectedEnergyExpression}
\\
0 &= \langle \mu | P \, \left(H - E\right) \, \mathrm{e}^{U} |\phi\rangle
\end{align}
\label{Eqns:PbarHbar}
\end{subequations}
where $|\mu\rangle$ is an excited determinant.  We have used the fact that $H$ and $P$ commute.

Alternatively, one could set up a similarity-transformed approach which requires similarity transformations of both the Hamiltonian and the projection operator.  Starting from the Schr\"odinger equation
\begin{equation}
P \, H \, \mathrm{e}^{U} |\phi\rangle = E \, P \, \mathrm{e}^{U} |\phi\rangle
\end{equation}
we multiply on the left by $\exp(-U)$ to obtain
\begin{equation}
\mathrm{e}^{-U} \, P \, H \, \mathrm{e}^{U} |\phi\rangle = E \,\mathrm{e}^{-U} \, P \, \mathrm{e}^{U} |\phi\rangle
\end{equation}
or equivalently,
\begin{equation}
\bar{P} \, \bar{H} |\phi\rangle = E \, \bar{P} |\phi\rangle.
\end{equation}
Recall that $\bar{H}$ is the similarity-transformed Hamiltonian (\textit{c.f.} Eqn. \ref{Eqn:DefBar}) while $\bar{P}$ is the similarity-transformed projection operator.  The energy and the amplitudes defining $U$ are then
\begin{subequations}
\begin{align}
E &= \frac{\langle \phi| \bar{P} \, \bar{H} |\phi\rangle}{\langle \phi| \bar{P} |\phi\rangle},
\\
0 &= \langle \mu | \bar{P} \, \left(\bar{H} - E\right) |\phi\rangle.
\end{align}
\label{Eqns:PbarHbar2}
\end{subequations}
Furthermore, by parameterizing the left hand state as in traditional coupled cluster theory, one can introduce the PCC analogue of the traditional coupled cluster energy functional, in terms of which the energy and amplitude equations are
\begin{subequations}
\begin{align}
\mathcal{E}
 &= \frac{\langle \phi| \left(1 + Z\right) \, \bar{P} \, \bar{H} |\phi\rangle}{\langle \phi| \left(1 + Z\right) \, \bar{P} |\phi\rangle},
\label{Eqn:LRPCCEnergy}
\\
0 &= \frac{\partial \mathcal{E}}{\partial Z_\mu},
\\
0 &= \frac{\partial \mathcal{E}}{\partial U_\mu}.
\end{align}
\label{Eqns:PbarHbarR}
\end{subequations}
Just as in traditional coupled cluster, one can show that all three formulations are equivalent, though it is important to note that, also as in traditional coupled cluster, we have $\mathcal{E} = E$ only when the cluster operator $U$ satisfies the amplitude equations.

Traditional coupled cluster theory finds it convenient to work with the similarity-transformed approach because the similarity transformation of the Hamiltonian naturally terminates at $\mathcal{O}(U^4)$, a consequence of the Hamiltonian being a two-body operator.  The projection operator, however, is an $N$-body operator, and its similiarity transformation therefore does not terminate in a convenient way.  Accordingly, we find it simpler to work in the formalism of Eqn. \ref{Eqns:PbarHbar}.  Of course one could truncate the commutator expansion of $\bar{P}$ manually, though so far we have considered only the approximation $\bar{P} \approx P$.  This approximation leads to results of significantly lower quality, so we will not discuss them.

While the similarity transformation of $P$ is cumbersome and essentially precludes the use of the similarity-transformed approach as it is traditionally expressed, we will find the need to evaluate the biorthogonal energy $\mathcal{E}$ from time to time.  To facilitate this, we define
\begin{equation}
\langle \phi| \left(1 + Z\right) \, \mathrm{e}^{-U} = \langle \phi| \tilde{Z}
\end{equation}
where $\tilde{Z}$ is a modified excitation operator when acting to the left and contains a component which creates no excitations.  This is essentially a consequence of Eqn. \ref{Eqn:SimTrans2NonSimTrans}.  To give a concrete example, when $U$ and $Z$ contain only single and double excitation operators, we may write
\begin{subequations}
\begin{align}
\tilde{Z}_2 &= Z_2,
\\
\tilde{Z}_1 &= Z_1 - \left(Z_2 \, U_1\right)_\mathrm{cd},
\\
\tilde{Z}_0 &= 1 - \left(Z_1 \, U_1\right)_\mathrm{cs} - \left(Z_2 \, U_2\right)_\mathrm{cs} + \frac{1}{2} \, \left(Z_2 \, U_1^2\right)_\mathrm{cs},
\end{align}
\end{subequations}
where the subscripts ``$\mathrm{cd}$'' and ``$\mathrm{cs}$'' respectively mean the connected deexcitation part and the connected scalar part.  In terms of $\tilde{Z}$ we may write the biorthogonal energy as
\begin{equation}
\mathcal{E} = \frac{\langle \phi| \tilde{Z} \, P \, H \, \mathrm{e}^{U} |\phi\rangle}{\langle \phi| \tilde{Z} \, P \, \mathrm{e}^{U} |\phi\rangle} = \frac{\sum_\nu \tilde{Z}_\nu \, \langle \nu| P \, H \, \mathrm{e}^{U} |\phi\rangle}{\sum_\nu \tilde{Z}_\nu \, \langle \nu| P \, \mathrm{e}^{U} |\phi\rangle},
\end{equation}
where $\tilde{Z}_\nu$ are the amplitudes defining the operator $\tilde{Z}$ and where the summation over $\nu$ includes the reference determinant $\langle \phi|$ as well as excited determinants.

\begin{figure*}[t]
\includegraphics[width=0.96\columnwidth]{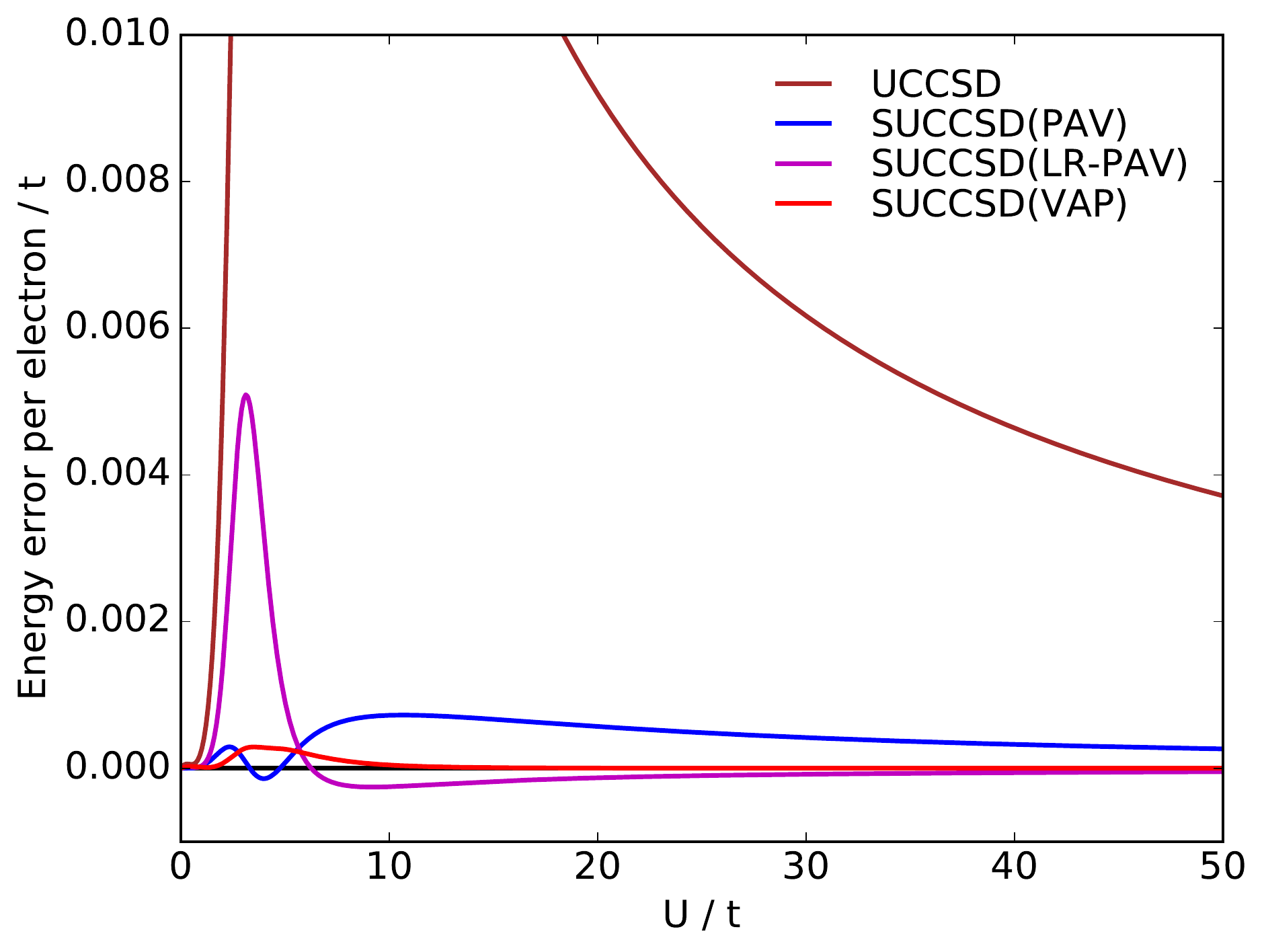}
\hfill
\includegraphics[width=0.96\columnwidth]{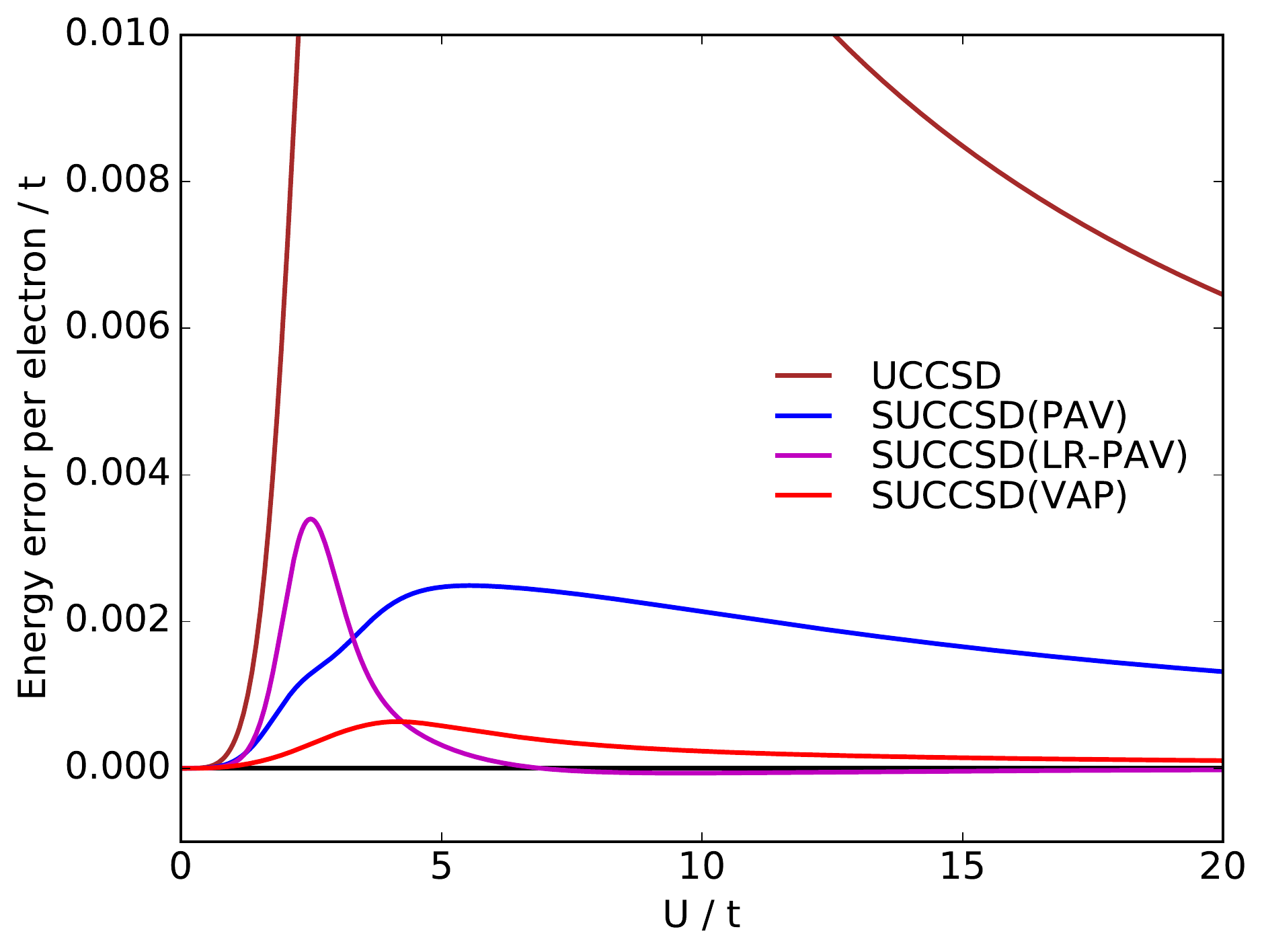}
\caption{Errors per electron with respect to the exact result in periodic half-filled Hubbard lattices. Left panel: 6-site Hubbard model. Right panel: 10-site Hubbard model. We use the SUHF broken symmetry determinant as a reference.
\label{Fig:PUCCFull}}
\end{figure*}

Our basic formulation is agnostic as to the precise way in which the symmetry projection is carried out.  In practice, we find it most convenient to use the integral representation outlined in Sec. \ref{Sec:SymmetryProjection}.  To do so, we need only define the requisite integral kernels, which are
\begin{subequations}
\begin{align}
\mathcal{N}(\Omega) &= \langle\phi| R(\Omega) \, \mathrm{e}^{U} |\phi\rangle,
\\
\mathcal{H}(\Omega) &= \frac{\langle\phi| R(\Omega) \, H \, \mathrm{e}^{U} |\phi\rangle}{\mathcal{N}(\Omega)},
\\
\mathcal{N}_{\mu}(\Omega) &= \frac{\langle\mu| R(\Omega) \, \mathrm{e}^{U} |\phi\rangle}{\mathcal{N}(\Omega)},
\\
\mathcal{H}_{\mu}(\Omega) &= \frac{\langle\mu| R(\Omega) \, H \, \mathrm{e}^{U} |\phi\rangle}{\mathcal{N}(\Omega)}.
\end{align}
\end{subequations}
From these kernels, we can obtain all quantities needed to evaluate the projected energy and amplitude equations:
\begin{subequations}
\begin{align}
\langle \phi| P \, \mathrm{e}^{U} |\phi\rangle &= \frac{1}{V_\Omega} \, \int \mathrm{d}\Omega \, w(\Omega) \, \mathcal{N}(\Omega),
\\
\langle \phi| P \, H \, \mathrm{e}^{U} |\phi\rangle &= \frac{1}{V_\Omega} \, \int \mathrm{d}\Omega \, w(\Omega) \, \mathcal{H}(\Omega) \, \mathcal{N}(\Omega),
\\
\langle \mu| P \, \mathrm{e}^{U} |\phi\rangle &= \frac{1}{V_\Omega} \, \int \mathrm{d}\Omega \, w(\Omega) \, \mathcal{N}_\mu(\Omega) \, \mathcal{N}(\Omega),
\\
\langle \mu| P \, H \, \mathrm{e}^{U} |\phi\rangle &= \frac{1}{V_\Omega} \, \int \mathrm{d}\Omega \, w(\Omega) \, \mathcal{H}_\mu(\Omega) \, \mathcal{N}(\Omega).
\end{align}
\label{Eqn:Kern2}
\end{subequations}

We should make an important caveat at this point.  In traditional coupled cluster theory the exponential can be truncated in a natural way.  Thus, for example, the CCSD equations require only up to the quadruple excitation part of $\exp(U)$.  This is so because we consider states $\langle \mu|$ which are no more than doubly-excited, and the Hamiltonian (as a two-body operator) cannot connect a doubly-excited bra state with more than a quadruply-excited ket state.  Unfortunately, in projected coupled cluster the exponential cannot be so conveniently truncated, due to the presence of the rotation operator $R(\Omega)$.  If we imagine defining rotated states 
\begin{equation}
\langle \mu(\Omega)| = \langle \mu| R(\Omega)
\end{equation}
then even when $\langle \mu| = \langle \mu(\Omega = 0)|$ is doubly-excited with respect to the ground state, $\langle \mu(\Omega \neq 0)|$ may have components which are very highly excited indeed.  Section \ref{Sec:Approximations} discusses how we circumvent this difficulty.

\subsection{Results}
Thus far we have been as general as possible.  Our PCC theory can be combined with any symmetry projection, and if the cluster operator is not truncated it provides the exact ground state wave function.  Here, we provide results for the special case of spin-projected unrestricted coupled cluster with single and double excitations (SUCCSD), and consider only projection onto a spin singlet.  We have adapted a full configuration interaction code to do SUCCSD calculations, which limits our exact SUCCSD to small systems.  A production version of the theory requires additional approximations which we introduce in Sec. \ref{Sec:Approximations} and which allow for a computationally efficient implementation; our exact results here thus provide a benchmark to compare against.  To ensure accuracy, we have used exact grids to integrate the various kernels where we have as many grid points as basis functions.  In practice the number of grid points needed for spin-projected UHF (SUHF) scales roughly as the square root of the number of basis functions and we expect fewer grid points are needed for SUCCSD.

In this section we also wish to address the importance of obtaining the amplitudes $U_\mu$ and potentially $Z_\mu$ from our PCC equations rather than from traditional coupled cluster.  In analogy with the literature on projected Hartree-Fock theory, we will refer to a variation after projection (VAP) approach in which the amplitudes solve the SUCCSD equations, where a projection after variation (PAV) approach means that we first solve the traditional unrestricted CCSD (UCCSD) equations and simply evaluate the projected energy.  Without further qualification, by the PAV energy we mean the energy expression of Eqn. \ref{Eqn:ProjectedEnergyExpression}, but we may also refer to the linear response PAV (LR-PAV) in which we use the biorthogonal expectation value of Eqn. \ref{Eqn:LRPCCEnergy}.  Recall that these two energy expressions yield different results when using UCCSD amplitudes but the same result if one solves for the $U$ amplitudes in the presence of the projection operator.  In other words, the VAP formulation of SUCCSD yields the same energy from the two different energy formulae.

\begin{figure*}[t]
\includegraphics[height=0.25\textheight]{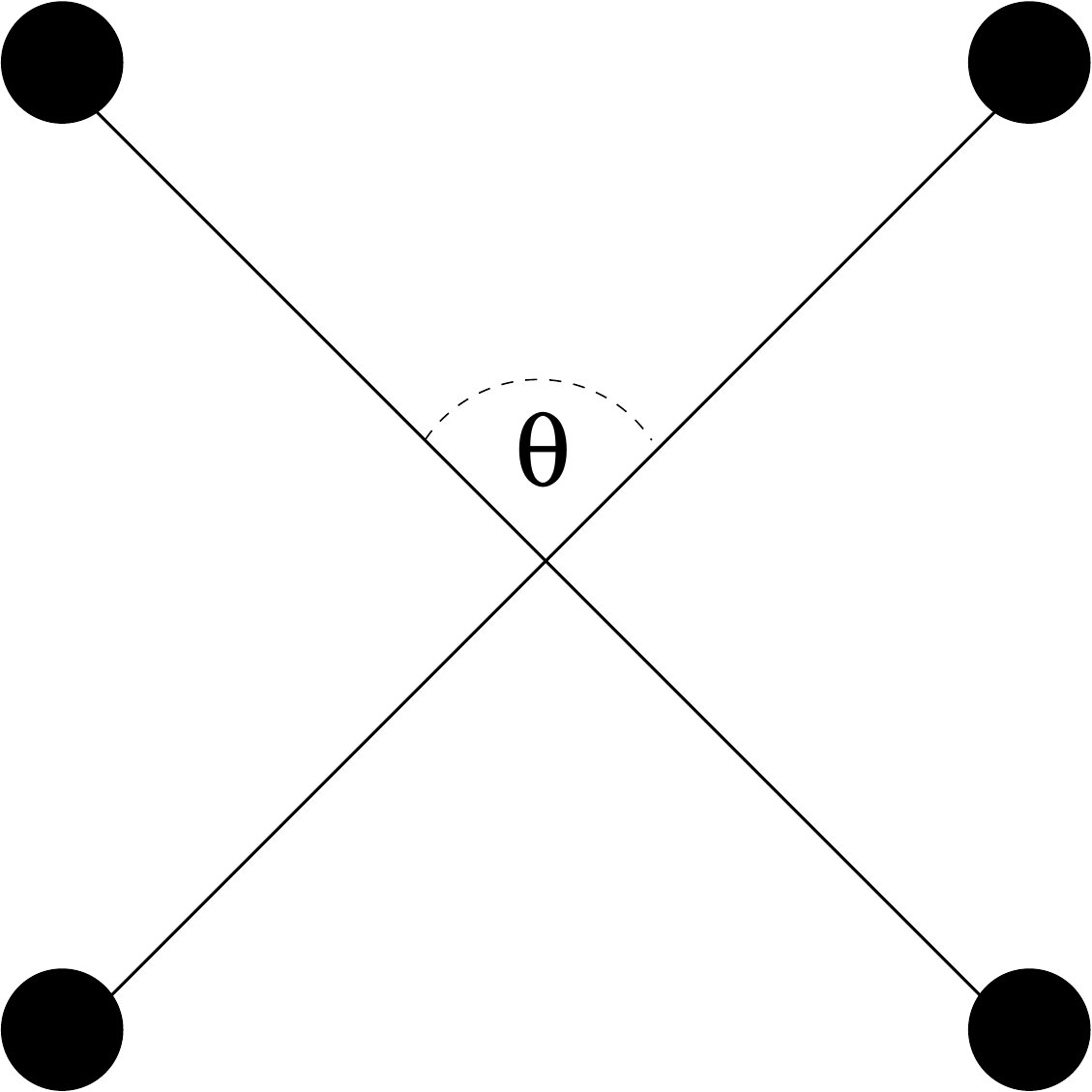}
\hfill
\includegraphics[height=0.28\textheight]{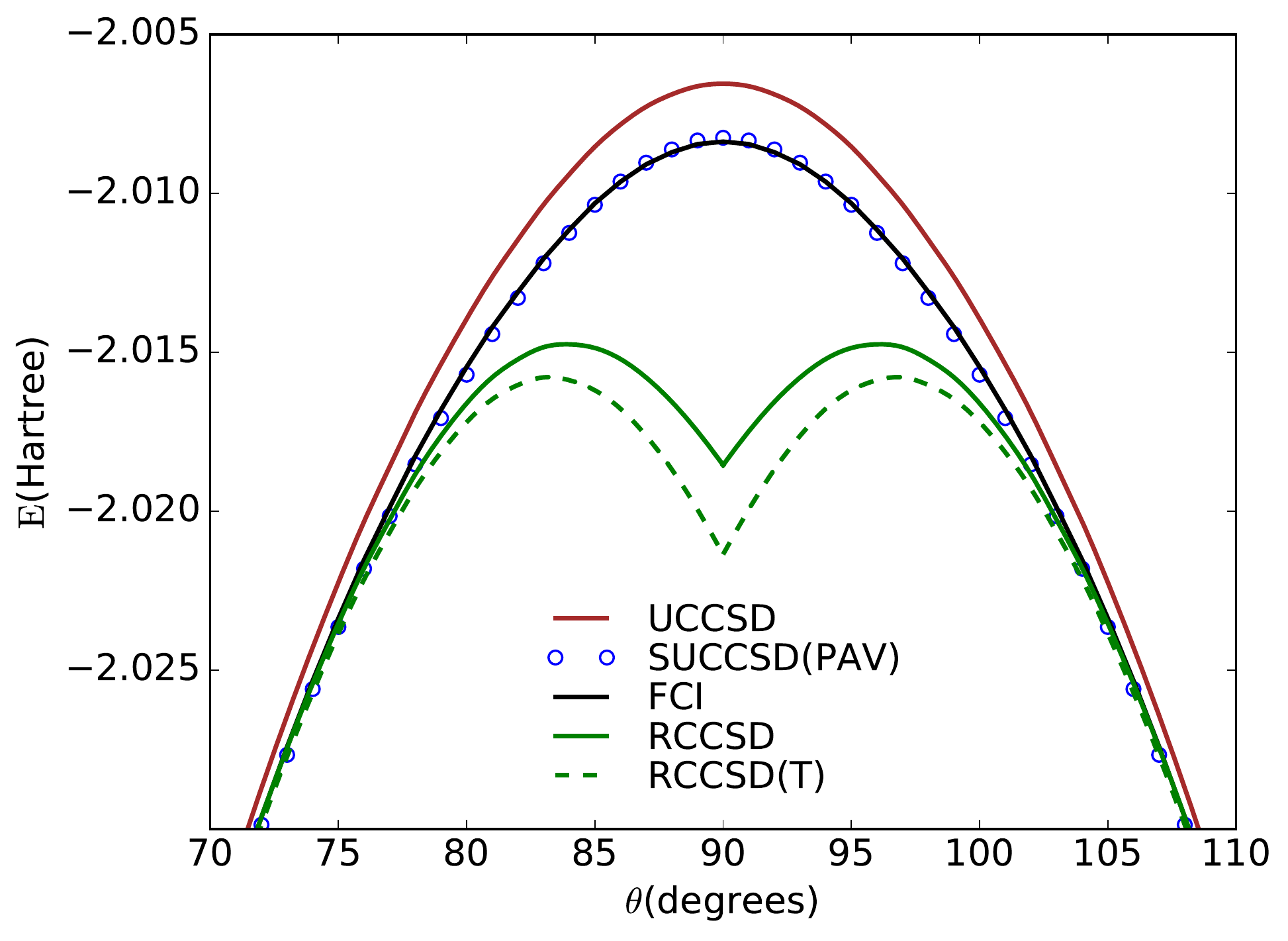}
\caption{Left panel: Schematic representation of the H$_4$ ring.  Right panel: Total energies for the H$_4$ ring as a function of angle.  While  restricted coupled cluster has a cusp, unrestricted coupled cluster and SUCCSD are smooth.  We use the UHF determinant as a reference and the cc-pVDZ basis set.  The left panel is reproduced with permission from Ref. \onlinecite{Qiu2017}.
\label{Fig:H4}}
\end{figure*}

Bearing all this in mind, let us first consider results in the Hubbard model Hamiltonian.\cite{Hubbard1963}  The Hubbard model describes electrons on a lattice, and the Hamiltonian is simply
\begin{equation}
H = -t \sum_{\langle \mu\nu \rangle} \left(c_{\mu_\uparrow}^\dagger \, c_{\nu_\uparrow} + c_{\mu_\downarrow}^\dagger \, c_{\nu_\downarrow}\right) + U \sum_\mu c_{\mu_\uparrow}^\dagger \, c_{\mu_\downarrow}^\dagger \, c_{\mu_\downarrow} \, c_{\mu_\uparrow}
\end{equation}
where $\mu$ and $\nu$ are lattice sites and the notation $\langle \mu\nu\rangle$ means that we include only sites connected in the lattice.   As the relative interaction strength $U/t$ increases, the system becomes more and more strongly correlated.  Our calculations use a one-dimensional lattice in which adjacent sites are connected.  We impose periodic boundary conditions, which is of more general interest and for which exact results are readily available through the Lieb-Wu algorithm.\cite{LiebWu}

Figure \ref{Fig:PUCCFull} shows results for the half-filled lattice with six and ten sites.  We obtain the reference determinant $|\phi\rangle$ from a variation after projection spin-projected UHF (SUHF) calculation, so that $|\phi\rangle$ is symmetry broken for all $U/t$.  One can see that the spin projection improves significantly upon UCCSD, regardless of whether the amplitudes solve the SUCCSD equations (VAP) or the UCCSD equations (PAV).  Unsurprisingly, our overall best results are obtained from the VAP approach.  In the strongly correlated limit, VAP and LR-PAV are comparable, and LR-PAV is in general superior to PAV without including the $Z$ amplitudes.  We do not know why the LR-PAV results have a relatively pronounced peak for small $U/t$.

Let us now turn to a few molecular examples. The first example is four hydrogen atoms placed on a circle of radius 3.284 Bohr,\cite{Troy2000} as depicted in Fig. \ref{Fig:H4}.  As the bond angle $\theta$ approaches $90^\circ$, the system becomes more strongly correlated as the ground and first excited states become nearly degenerate (and exactly degenerate precisely at the high symmetry point).  Where restricted CCSD (RCCSD) has a cusp and incorrectly predicts the existence of a minimum at $\theta = 90^\circ$, UCCSD remedies these failures at the cost of correct spin symmetry.  Moreover, UCCSD undercorrelates somewhat. In contrast, SUCCSD is virtually atop the exact result, though we have been able to compute only the PAV result because the cc-pVDZ basis set which we use is too large for our full configuration interaction code.  Incorporating linear response adds virtually nothing in this case.

\begin{figure*}[t]
\includegraphics[width=0.96\columnwidth]{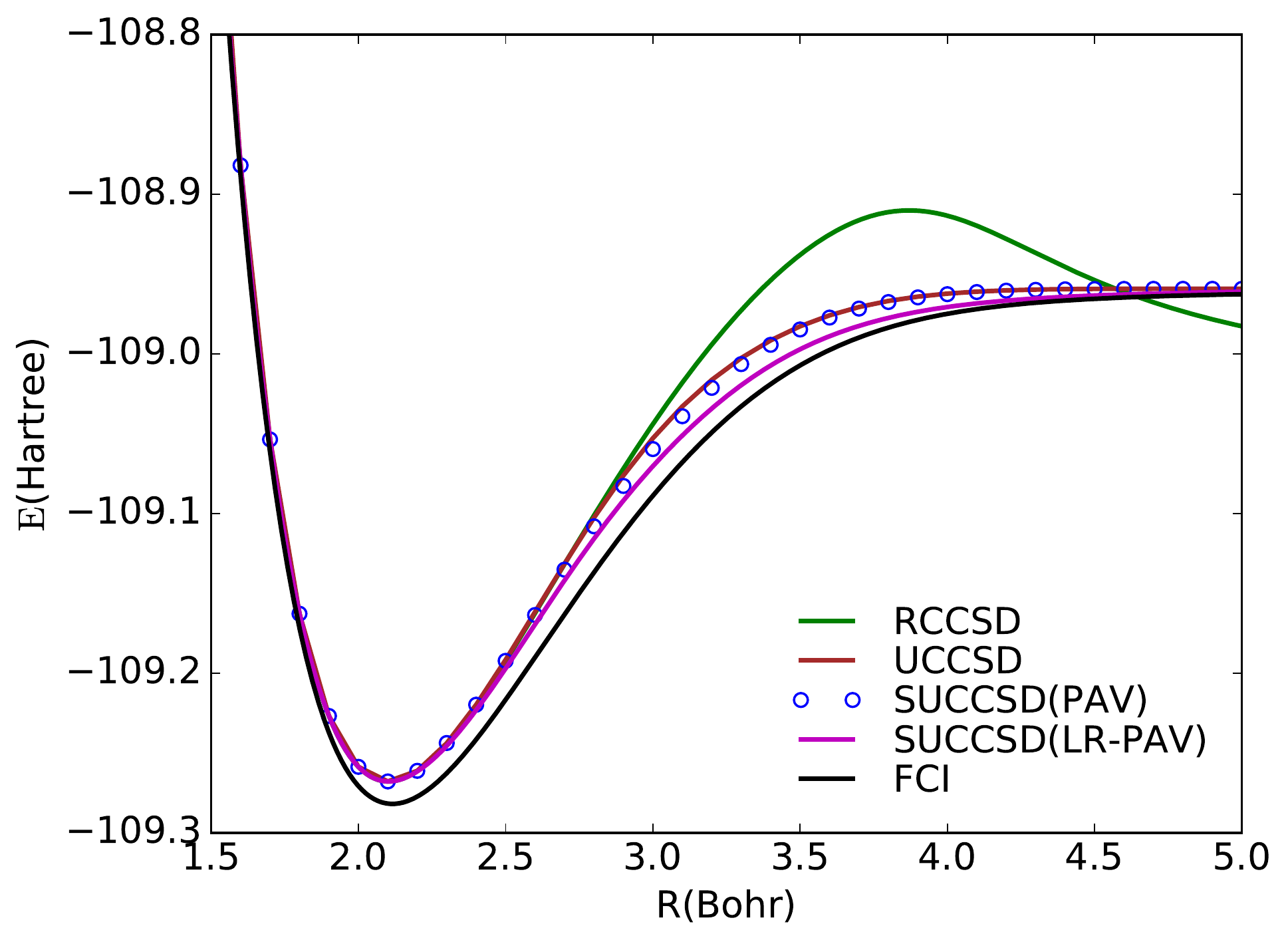}
\hfill
\includegraphics[width=0.96\columnwidth]{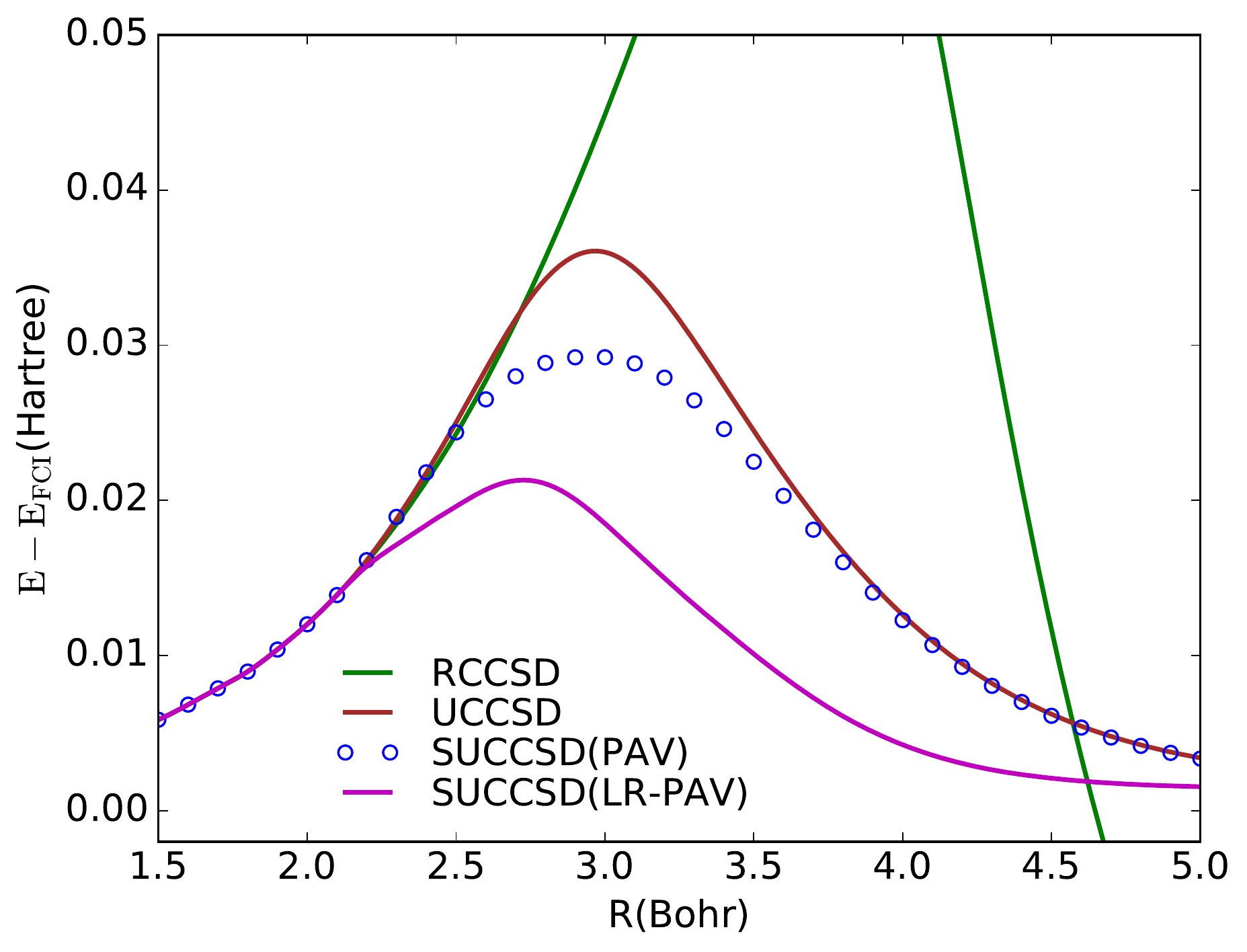}
\caption{Energies of the N$_2$ molecule using the cc-pVDZ basis set. Left panel: Total energies. Right panel: Errors with respect to full configuration interaction (FCI).
\label{Fig:N2}}
\end{figure*}

Finally we consider the dissociation of N$_2$ in the cc-pVDZ bas set, as shown in Fig. \ref{Fig:N2}.  Restricted CCSD overcorrelates badly at the dissociation limit and has an unphysical bump in the potential energy curve.  Adding perturbative triples to obtain RCCSD(T) makes the situation even worse (not shown).  If one uses an unrestricted reference the results are much improved.  Results are better yet with SUCCSD, though again we have not been able to carry out the VAP calculations.  For small $R$, the UHF reference does not break symmetry, so symmetry projection has no effect.  This causes the break in the LR-PAV curve.  The LR-PAV curve is smoothed by using the SUHF reference instead (not shown), though overall the reference-dependence is weak.

We should make one small caveat about our N$_2$ results.  Because the system was beyond the scope of our full configuration interaction code, we were unable to evaluate the Hamiltonian and norm kernels exactly.  Accordingly, we truncated the exponential of the cluster operator in both and retained terms only through $\mathcal{O}(U_2^3)$.  While exact results would require us to keep terms through $\mathcal{O}(U_2^7)$, we have found that the energy is in practice converged at the truncation we have shown, presumably because there are six strongly correlated electrons.  We have kept the $U_1$ terms to infinite order, as they can be absorbed into transforming the Hamiltonian, discussed in detail in Appendix \ref{Appendix:U1}.

Nonetheless, the inability to exactly compute the Hamiltonian and norm kernels is a serious limitation, and simply truncating the exponential is not always practical.  Accordingly, we now introduce our disentangled cluster formalism, which allows us to circumvent this limitation.

\section{The Disentangled Cluster Formalism
\label{Sec:Approximations}}
As we have seen, the chief difficulty in evaluating the kernels needed to do PCC calculations is that in principle the exponential must be expanded to all powers of $U$, which is not generally feasible.  We now turn our attention to overcoming this obstacle.  We will need to make some preliminary simplifications first.

We begin by replacing the rotation operator, which in general is a unitary Thouless transformation,\cite{Thouless1960} with a particle-hole-style Thouless transformation applied to the bra:
\begin{equation}
\langle\phi| R(\Omega) = \langle\phi| R(\Omega) |\phi\rangle \; \langle\phi| \mathrm{e}^{V_1(\Omega)}
\end{equation}
where $V_1(\Omega)$ is a single de-excitation operator.  Having done so, the norm and Hamiltonian kernels become
\begin{subequations}
\begin{align}
\mathcal{N}(\Omega) &= \langle \phi|R(\Omega)|\phi\rangle \, \langle \phi| \mathrm{e}^{V_1(\Omega)} \, \mathrm{e}^{U} |\phi\rangle,
\\
\mathcal{H}(\Omega) &= \frac{\langle \phi| \bar{H}_{V_1}(\Omega) \, \mathrm{e}^{V_1(\Omega)} \, \mathrm{e}^{U} |\phi\rangle}{\langle \phi| \mathrm{e}^{V_1(\Omega)} \, \mathrm{e}^{U} |\phi\rangle},
\end{align}
\end{subequations}
where we have introduced the similarity-transformed Hamiltonian
\begin{equation}
\bar{H}_{V_1}(\Omega) = \mathrm{e}^{V_1(\Omega)} \, H \, \mathrm{e}^{-V_1(\Omega)}.
\end{equation}
Since $V_1(\Omega)$ is a one-body operator, $\bar{H}_{V_1}(\Omega)$ contains no more than two-body operators (assuming that $H$ itself had only one- and two-body operators).  Henceforth we will omit the explicit $\Omega$-dependence to reduce notational clutter.

In addition to the ground-state reduced kernels, we also need excited kernels $\mathcal{N}_\mu$ and $\mathcal{H}_\mu$.  To obtain these, we first write
\begin{equation}
\langle \mu|  = \langle \phi| \mathcal{Q}_\mu
\end{equation}
where $\mathcal{Q}_\mu$ is an excitation operator when acting to the left.  Having done so, we can then use
\begin{subequations}
\begin{align}
\langle \mu| R &= \langle \phi| R \, R^{-1} \, \mathcal{Q}_\mu \, R
\\
 &= \langle \phi|R|\phi\rangle \, \langle \phi| \mathrm{e}^{V_1} \, R^{-1} \, \mathcal{Q}_\mu \, R \, \mathrm{e}^{-V_1} \, \mathrm{e}^{V_1}
\\
 &= \langle \phi|R|\phi\rangle \, \langle \phi| \tilde{\mathcal{Q}}_\mu \, \mathrm{e}^{V_1}
\end{align}
\end{subequations}
where
\begin{equation}
\tilde{\mathcal{Q}}_\mu = \mathrm{e}^{V_1} \, R^{-1} \, \mathcal{Q}_\mu \, R \, \mathrm{e}^{-V_1}.
\end{equation}
The excited norm and Hamiltonian kernels are thus
\begin{subequations}
\begin{align}
\mathcal{N}_\mu &= \frac{\langle \phi|\tilde{\mathcal{Q}}_\mu \, \mathrm{e}^{V_1} \, \mathrm{e}^{U} |\phi\rangle}{\langle \phi| \mathrm{e}^{V_1} \, \mathrm{e}^{U} |\phi\rangle},
\\
\mathcal{H}_\mu &= \frac{\langle \phi| \tilde{\mathcal{Q}}_\mu \, \bar{H}_{V_1} \, \mathrm{e}^{V_1} \, \mathrm{e}^{U} |\phi\rangle}{\langle \phi| \mathrm{e}^{V_1} \, \mathrm{e}^{U} |\phi\rangle}.
\end{align}
\end{subequations}

We can evaluate the action of $\tilde{\mathcal{Q}}_\mu$ straightforwardly, because $\exp(V_1) \, R^{-1}$ is just an orbital transformation operator, given in the molecular orbital basis as
\begin{equation}
\mathbf{T} = \exp\begin{pmatrix}\mathbf{0} & \mathbf{V}_\mathrm{ov} \\ \mathbf{0} & \mathbf{0}\end{pmatrix} \, \mathbf{R}^{-1} = \begin{pmatrix} \mathbf{1} & \mathbf{V}_\mathrm{ov} \\ \mathbf{0} & \mathbf{1} \end{pmatrix} \, \mathbf{R}^{-1},
\label{Eqn:DefTMat}
\end{equation}
where $\mathbf{V}_\mathrm{ov}$ contains the amplitudes $v^i_a$ defining $V_1$ and where $\mathbf{R}$ is the matrix representation of the rotation operator.  For example, the rotation operator $R$ for unrestricted spin projection is $\exp(-\mathrm{i} \, \beta \, S_y)$ so the matrix $\mathbf{R}^{-1}$ is $\exp(\mathrm{i} \, \beta \, \mathbf{S}_y)$ where $\mathbf{S}_y$ is the matrix representation of the $S_y$ operator.

As concrete examples, the singly- and doubly-excited kernels are 
\begin{subequations}
\begin{align}
\mathcal{N}_i^a
 &= \frac{\langle \phi_i^a| R \, \mathrm{e}^U|\phi\rangle}{\mathcal{N}}
  = (T^{-1})^a_p \, T_i^q \, \tilde{N}^p_q,
\\
\mathcal{N}_{ij}^{ab}
 &= \frac{\langle \phi_{ij}^{ab}| R \, \mathrm{e}^U|\phi\rangle}{\mathcal{N}}
  = (T^{-1})^a_p \, (T^{-1})^b_r \, T^q_i \, T_j^s \, \tilde{N}_{qs}^{pr},
\\
\mathcal{H}_i^a
 &= \frac{\langle \phi_i^a| R \, H \,  \mathrm{e}^U|\phi\rangle}{\mathcal{N}}
  = (T^{-1})^a_p \, T_i^q \, \tilde{H}^p_q,
\label{Eqn:DefHia}
\\
\mathcal{H}_{ij}^{ab}
 &= \frac{\langle \phi_{ij}^{ab}|R \, H \,  \mathrm{e}^U|\phi\rangle}{\mathcal{N}}
  = (T^{-1})^a_p \, (T^{-1})^b_r \, T^q_i \, T_j^s \, \tilde{H}_{qs}^{pr}.
\end{align}
\end{subequations}

We must make several comments to clarify the foregoing.  First, we have followed the usual notation that indices $i$ and $j$ denote orbitals occupied in $|\phi\rangle$, while $a$ and $b$ denote orbitals unoccupied in $|\phi\rangle$ and $p$, $q$, $r$, and $s$ are general.  Determinants $\langle \phi_i^a|$ and $\langle \phi_{ij}^{ab}|$ are singly- and doubly-excited, respectively.  Our convention for matrices is that row (column) indices of the matrix $\mathbf{T}$ correspond to upper (lower) indices of $T_p^q$.  Finally, we have introduced auxiliary quantities such as $\tilde{\mathcal{N}}_p^q$.  These are
\begin{subequations}
\begin{align}
\tilde{\mathcal{N}}^p_q 
 &= \frac{\langle\phi| c_q^\dagger \, c_p \, \mathrm{e}^{V_1} \, \mathrm{e}^U|\phi\rangle}{\mathcal{N}},
\\
\tilde{\mathcal{N}}^{pr}_{qs}
 &= \frac{\langle\phi| c_q^\dagger \, c_s^\dagger \, c_r \, c_p \, \mathrm{e}^{V_1} \, \mathrm{e}^U|\phi\rangle}{\mathcal{N}},
\\
\tilde{\mathcal{H}}^p_q 
 &= \frac{\langle\phi| c_q^\dagger \, c_p \, \bar{H}_{V_1} \, \mathrm{e}^{V_1} \, \mathrm{e}^U|\phi\rangle}{\mathcal{N}},
\\
\tilde{\mathcal{H}}^{pr}_{qs}
 &= \frac{\langle\phi| c_q^\dagger \, c_s^\dagger \, c_r \, c_p \, \bar{H}_{V_1} \, \mathrm{e}^{V_1} \, \mathrm{e}^U|\phi\rangle}{\mathcal{N}}.
\end{align}
\label{Eqn:AuxilliaryKernels}
\end{subequations}
These objects are sparse, and their non-zero elements can be worked out by considering the action of the fermionic creation and annihilation operators on the reference bra determinant $\langle \phi|$.  We include the non-zero components in Appendix \ref{Appendix:NonZeroNH}.

\subsection{The Disentangled Cluster Operator}
The remaining difficulty in evaluating the various kernels we need is handling the state $\mathrm{e}^{V_1} \, \mathrm{e}^{U}|\phi\rangle$.   If we cannot do so, all of the foregoing is in vain.  To accomplish this task, we now introduce our disentangled cluster approximation.  We will specialize to the case of coupled cluster doubles where $U = U_2$, with $U_2$ a double-excitation operator.  Singles need not be explicitly included because the single-excitation operator $U_1$ can be absorbed into transforming the Hamiltonian and symmetry operators, as we discuss in Appendix \ref{Appendix:U1}.  Higher excitations could be included in principle, and the basic framework generalizes to include them, but at increasing algebraic and computational complexity.

The basic idea is to rewrite $\mathrm{e}^{V_1} \, \mathrm{e}^{U_2} |\phi\rangle$ as an exponential of cluster operators, denoted by $W_k$, acting on $|\phi\rangle$.  That is, when acting to the right, $\mathrm{e}^{U_2}$ is an excitation operator and $\mathrm{e}^{V_1}$ is a deexcitation operator, and their product, acting on the reference determinant, can be expressed purely in terms of excitations together with a normalization constant.  One has simply
\begin{equation}
\mathrm{e}^{V_1} \, \mathrm{e}^{U_2}|\phi\rangle = \mathrm{e}^{W_0 + W_1 + W_2 + W_3 + \cdots} |\phi\rangle
\label{Eqn:WDef1}
\end{equation}
where
\begin{subequations}
\begin{align}
W_k &= \sum_n \frac{1}{(2n-k)!n!} \, \left(V_1^{2n-k} \, U_2^n\right)_\mathrm{ce}
\label{Eqn:WDef2}
\\
 &= \frac{1}{(k!)^2} \, \sum W_{i_1 \ldots i_k}^{a_1 \ldots a_k} \, c_{a_1}^\dagger \ldots c_{a_k}^\dagger \, c_{i_k} \, c_{i_1}.
\label{Eqn:WDef3}
\end{align}
\label{Eqn:WDef}
\end{subequations}
Here, the subscript ``$\mathrm{ce}$'' means only the connected excitation part is retained.  The sum over $n$ in defining $W_k$ must include all possible excitation levels.  Equation \ref{Eqn:WDef} is a rigorous result, proven in Appendix \ref{Appendix:ProveW}, and is central to the remainder of this paper.  We refer to it as the disentanglement equation, and to the $W_k$'s as the disentengled cluster operators.  Equations \ref{Eqn:WDef2} and \ref{Eqn:WDef3} define $W_k$ for $k >= 1$; $W_0$ is a number, given by
\begin{equation}
\mathrm{e}^{W_0} = \langle \phi|\mathrm{e}^{V_1} \, \mathrm{e}^{U_2} |\phi\rangle,
\end{equation}
and can be evaluated by our formula for $W_k$ with $k=0$ but replacing the connected excitation part with the connected scalar part.

Introducing the disentangled clusters makes it possible to work in an orthogonal framework.  The necessary kernels can all be evaluated with low excitation rank $W_k$:
\begin{subequations}
\begin{align}
\mathcal{N} &= \mathrm{e}^{W_0} \, \langle\phi| R |\phi\rangle
\\
\mathcal{H} &= \langle\phi| \bar{H}_{V_1} \, (W_1 + W_2+ \frac{1}{2}W_1^2 ) |\phi\rangle
\\
\tilde{\mathcal{N}}^a_i &= W^a_i,
\\
\tilde{\mathcal{N}}_{ij}^{ab} &= W_{ij}^{ab} + W_i^a \, W_j^b - W_j^a \, W_i^b,
\\                      
\tilde{\mathcal{H}}^a_i &= \langle \phi^a_i |\bar{H}_{V_1} \left(C_1 + C_2 + C_3\right) |\phi\rangle,
\\
\tilde{\mathcal{H}}^{ab}_{ij} &= \langle\phi^{ab}_{ij}| \bar{H}_{V_1} \left(C_1 + C_2 + C_3 + C_4\right)  |\phi\rangle.
\end{align}
\label{Eqn:KernFromUt2}
\end{subequations}
Here, $C_i$ is the $i$-fold excitation part of $\mathrm{e}^{W}$:
\begin{subequations}
\begin{align}
C_1 &= W_1,
\\
C_2 &= \frac{1}{2} \, W_1^2 + W_2,
\\
C_3 &= \frac{1}{3!}\, W_1^3 + W_1 \, W_2 + W_3,
\\
C_4 &= \frac{1}{4!}\, W_1^4 + \frac{1}{2} \, W_1^2 \, W_2 + \frac{1}{2} \, W_2^2 + W_1 \, W_3 + W_4.
\end{align}
\label{Def:Cn}
\end{subequations}
We need only up to $W_2$ to evaluate the energy without linear response.  Evaluating the energy with linear response or solving the amplitude equations requires up to $W_4$.

\subsection{Evaluating the Disentangled Cluster Operators}
At this point, we have everything we need for projected CCD or CCSD, provided only that we can obtain the disentangled cluster amplitudes defining $W$.  We will discuss several strategies for doing so.  We emphasize again that because $W$ depends on $V_1$ and $V_1$ depends on the integration variable $\Omega$, $W$ is $\Omega$ dependent, though we have suppressed this dependence for brevity.

\subsubsection{Truncated Disentangled Cluster Operators}
The simplest approach is to presume that $U_2$ is small and truncate the summation defining $W$.  If we truncate the sum over $U_2$ in Eqn. \ref{Eqn:WDef} to obtain amplitudes $W$ from which the kernels are extracted, however, we run into a problem: when the rotated state $\langle \phi|R$ has small overlap with $|\phi\rangle$, then $V_1$ is large and the series defining $W$ converges slowly, with terms having large amplitudes and alternating signs.  For example, in the half-filled Hubbard Hamiltonian for large $U/t$ the mean-field ground state has a N\'eel structure in which each site is occupied by one electron and the electrons on neighboring sites point in opposite directions.  Spin projection at $\beta = \pi$ amounts to flipping each spin, so $|\phi\rangle$ and $\langle \phi|R$ are strictly orthogonal; thus $V_1(\beta = \pi)$ cannot even be defined but the limit of $V_1$ as $\beta$ approaches $\pi$ is infinite.

We can do better by truncating the exponential itself.  For example, we could expand
\begin{equation}
e^{W_0} = 1 + (W_0^{(1)}) + \left(W_0^{(2)} + \frac{1}{2} \, W_0^{(1)} \, W_0^{(1)}\right) + \ldots
\end{equation}
where $W_0^{(n)}$ is the component of $W_0$ which has $n$ powers of $U_2$.  Note that except for the norm, all integrals in Eqn. \ref{Eqn:Kern2} are integrals of the product of the norm kernel and another kernel; in these cases, the product is truncated as a whole.  For example, we would write
\begin{equation}
\mathcal{H} \, \mathcal{N} = \mathcal{H}^{(0)} \, \mathcal{N}^{(0)} + \left(\mathcal{H}^{(1)} \, \mathcal{N}^{(0)} + \mathcal{H}^{(0)} \, \mathcal{N}^{(1)}\right) + \ldots
\end{equation}
where similarly $\mathcal{H}^{(n)}$ and $\mathcal{N}^{(n)}$ are portions of the kernel which contain $n$ powers of $U_2$.

The process of expressing a certain order of kernel in terms of $U_2$ and $V_1$ is tedious and error-prone. The derivation of equations is facilitated by an in-home algebra manipulator,\cite{Jinmo} and this part of code is generated by the accompanying automatic code generator.\cite{Jinmo} For this work, the energy functional without response is implemented up to fifth order in $U_2$, while the one with response is implemented up to third order.  For the energy functional without response, the energy can be evaluated through $U_2^3$ at $\mathcal{O}(N^6)$ cost, through $U_2^4$ at $\mathcal{O}(N^7)$ cost, and through $U_2^5$ at $\mathcal{O}(N^8)$ cost.  The energy functional with response can be evaluated up through $U_2^2$ at $\mathcal{O}(N^6)$ cost and through $U_2^3$ at $\mathcal{O}(N^8)$ cost.  All of these costs are per grid point in the numerical integration grid.

\subsubsection{Algebraic Equations for Disentangled Cluster Operators}
A second approach to obtaining the disentangled cluster operators generalizes work of Jeziorski,\cite{Jeziorski1993} that we rediscovered independently.  Let us define the virtual occupation operator as $n_{V} = \sum_{a} c^{\dagger}_a c_a $. It can be shown that for a $k$th level excitation $U_k$, or a $k$th level deexcitation $V_k$,
\begin{subequations}
\begin{align}
[n_V, U_k] &= k \, U_k,
\\
[n_V, V_k] &= -k \, V_k.
\end{align}
\end{subequations}
As a result,
\begin{subequations}
\begin{align}
[n_V, \mathrm{e}^{U_2}] &= 2 \, U_2 \, \mathrm{e}^{U_2},
\\
[n_V, \mathrm{e}^{V_1}] &= - V_1 \, \mathrm{e}^{V_1},
\\
[n_V, \mathrm{e}^{W}] &= \sum_k k \, W_k \, \mathrm{e}^{W}.
\end{align}
\end{subequations}

Applying these results to both sides of Eqn. \ref{Eqn:WDef} separately leads us to
\begin{subequations}
\begin{align}
n_V \, \mathrm{e}^{V_1} \, \mathrm{e}^{U_2}|\phi\rangle
  &= [n_V, \mathrm{e}^{V_1} \, \mathrm{e}^{U_2}] |\phi\rangle
\\
  &= (-V_1 \, \mathrm{e}^{V_1} \, \mathrm{e}^{U_2} + 2 \, \mathrm{e}^{V_1} \, U_2 \, \mathrm{e}^{U_2}) |\phi\rangle
\nonumber
\\
  &= (-V_1 + 2 \mathrm{e}^{V_1} \, U_2 \, \mathrm{e}^{-V_1}) \, \mathrm{e}^{V_1}\, \mathrm{e}^{U_2}|\phi\rangle
\nonumber
\\
  &= J \, \mathrm{e}^{W} |\phi\rangle,
\nonumber
\\
J &= -V_1 + 2 \, \mathrm{e}^{V_1} \, U_2 \, \mathrm{e}^{-V_1}.
\end{align}
\end{subequations}
On the other hand, 
\begin{equation}
n_V \, \mathrm{e}^{W} |\phi\rangle = [n_V, \mathrm{e}^{W}] |\phi\rangle = \sum k \, W_k \, \mathrm{e}^{W} |\phi\rangle.
\end{equation}

Together, these equations show that
\begin{equation}
J \, \mathrm{e}^W |\phi\rangle = \sum k \, W_k \, \mathrm{e}^W |\phi\rangle
\end{equation}
which implies
\begin{equation}
\sum k \, W_k |\phi\rangle = \mathrm{e}^{-W} \, J \, \mathrm{e}^W |\phi\rangle = \bar{J} |\phi\rangle.
\end{equation}
Amplitudes of $W_k$ can be extracted through left-multiplication with excited determinants.  For example
\begin{subequations}
\begin{align}
W^a_i &= \langle\phi^a_i| \bar{J} |\phi\rangle,
\\
W^{ab}_{ij} &= \frac{1}{2} \langle\phi^{ab}_{ij}| \bar{J} |\phi\rangle,
\end{align}
\label{Eqn:Alg}
\end{subequations}
or, more abstractly,
\begin{equation}
W_k = \frac{1}{k} \, \langle k|\bar{J} |\phi\rangle.
\label{Eqn:Alg2}
\end{equation}

Notice that $J$ contains one- and two-body parts only.  Thus, the expressions for $W$ look like the traditional coupled cluster equations, with $H$ replaced by $J$, and accordingly, $W_k$ is coupled with up to $W_{k+2}$.  Note also that because $W_0$ is simply a number and these algebraic equations are derived from commutators, we cannot use them to evaluate or even approximate $W_0$.

\begin{figure*}[t]
\includegraphics[width=0.96\columnwidth]{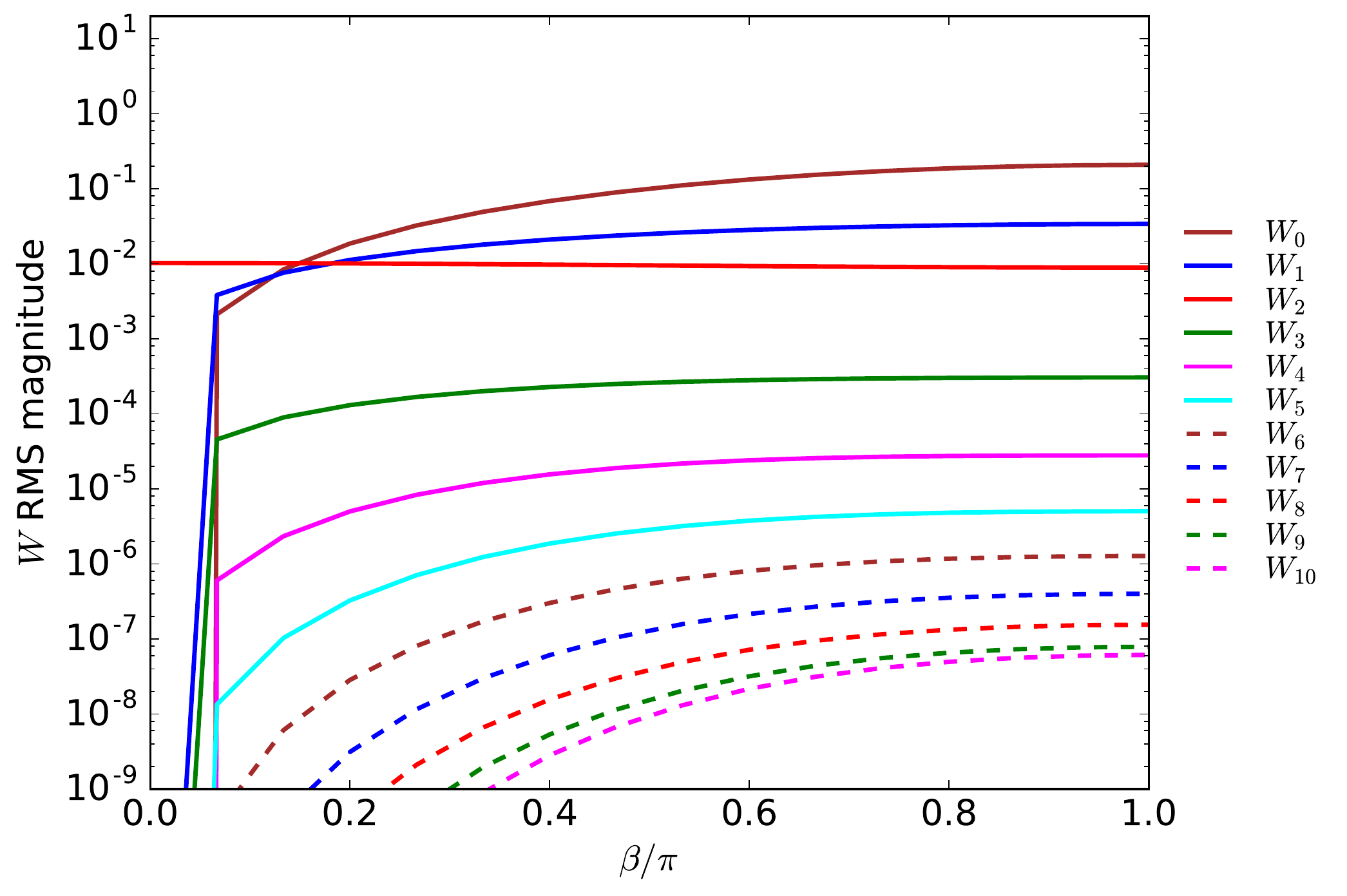}
\hfill
\includegraphics[width=0.96\columnwidth]{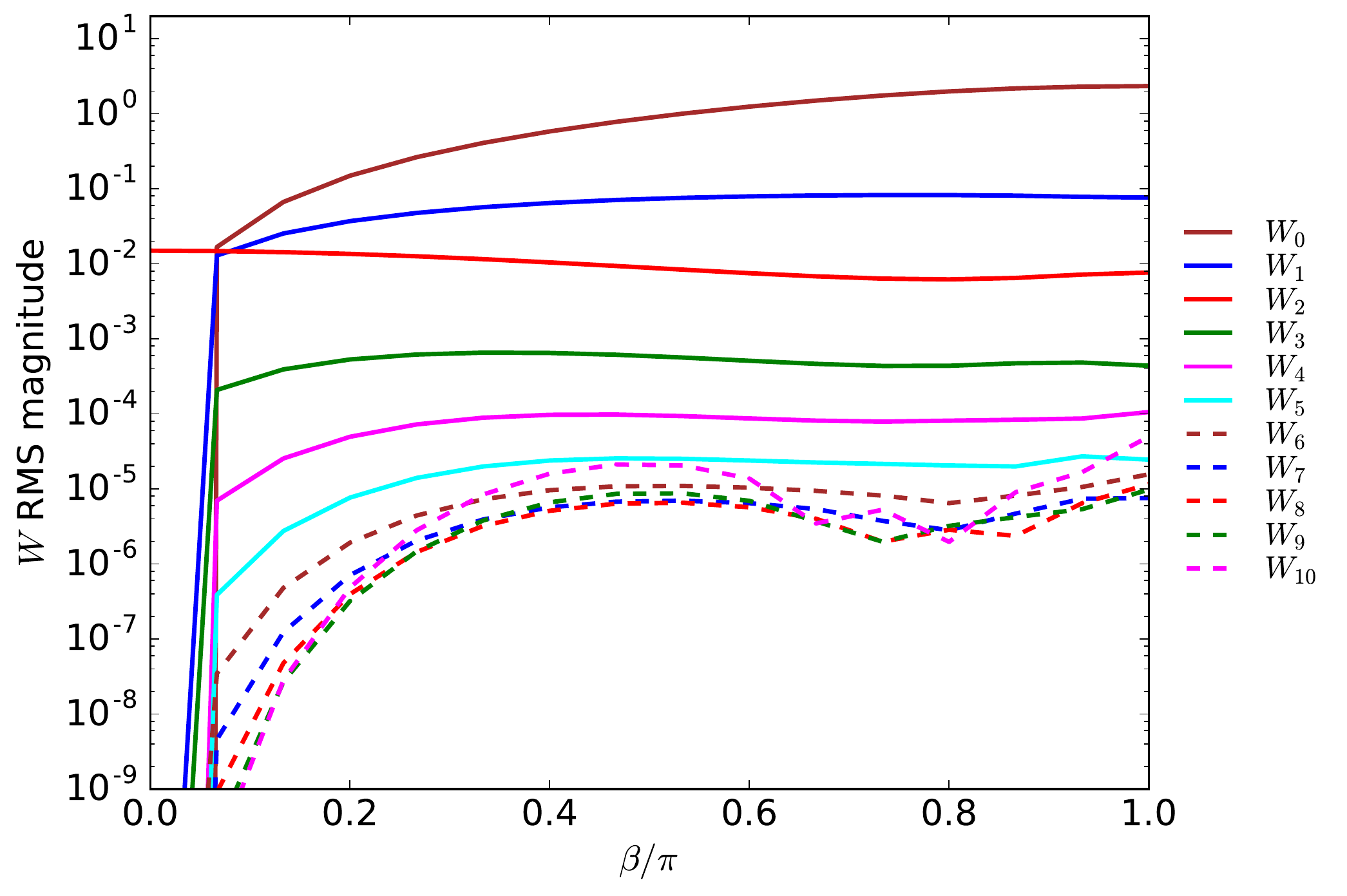}
\\
\includegraphics[width=0.96\columnwidth]{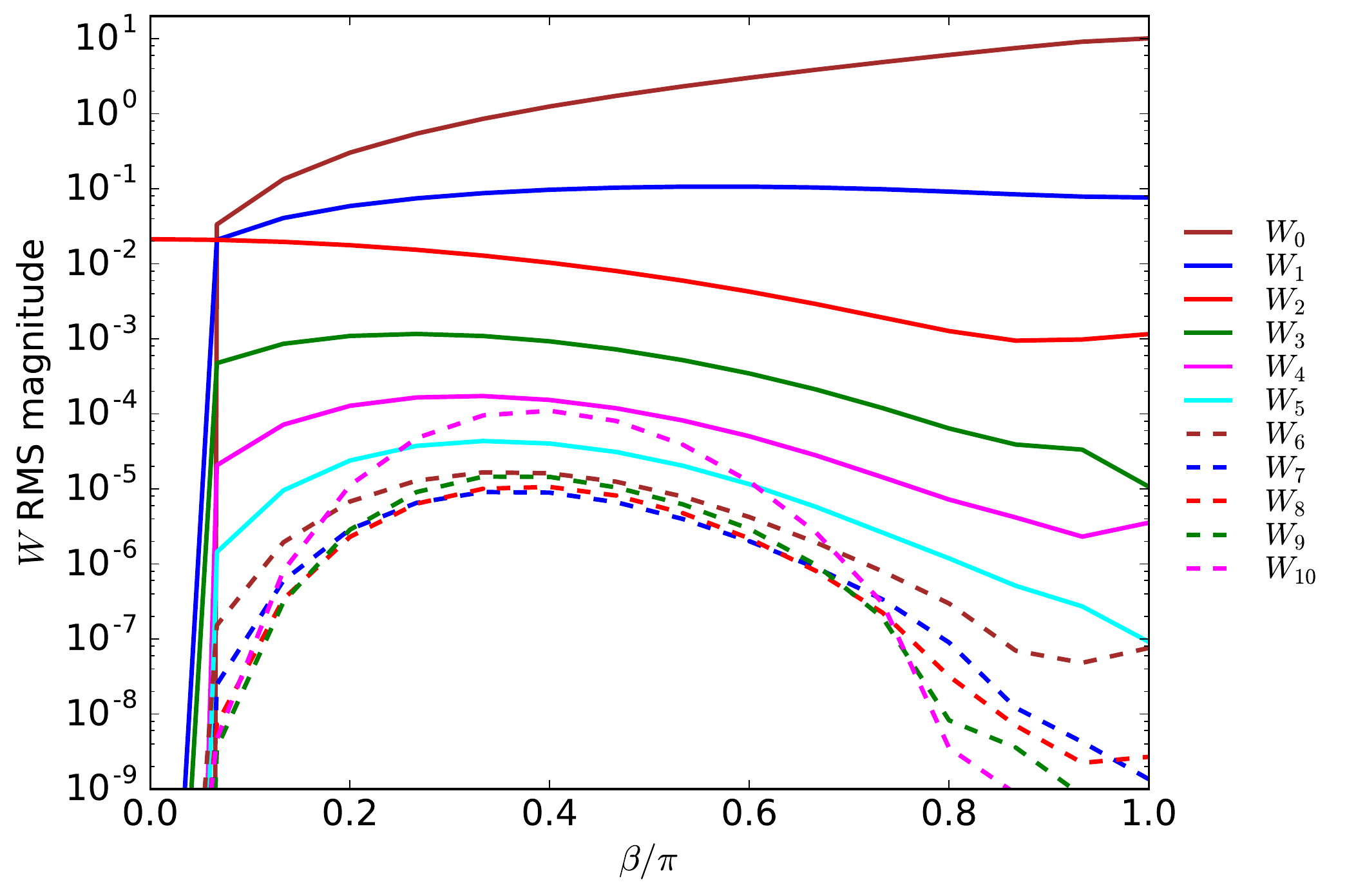} 
\hfill
\includegraphics[width=0.96\columnwidth]{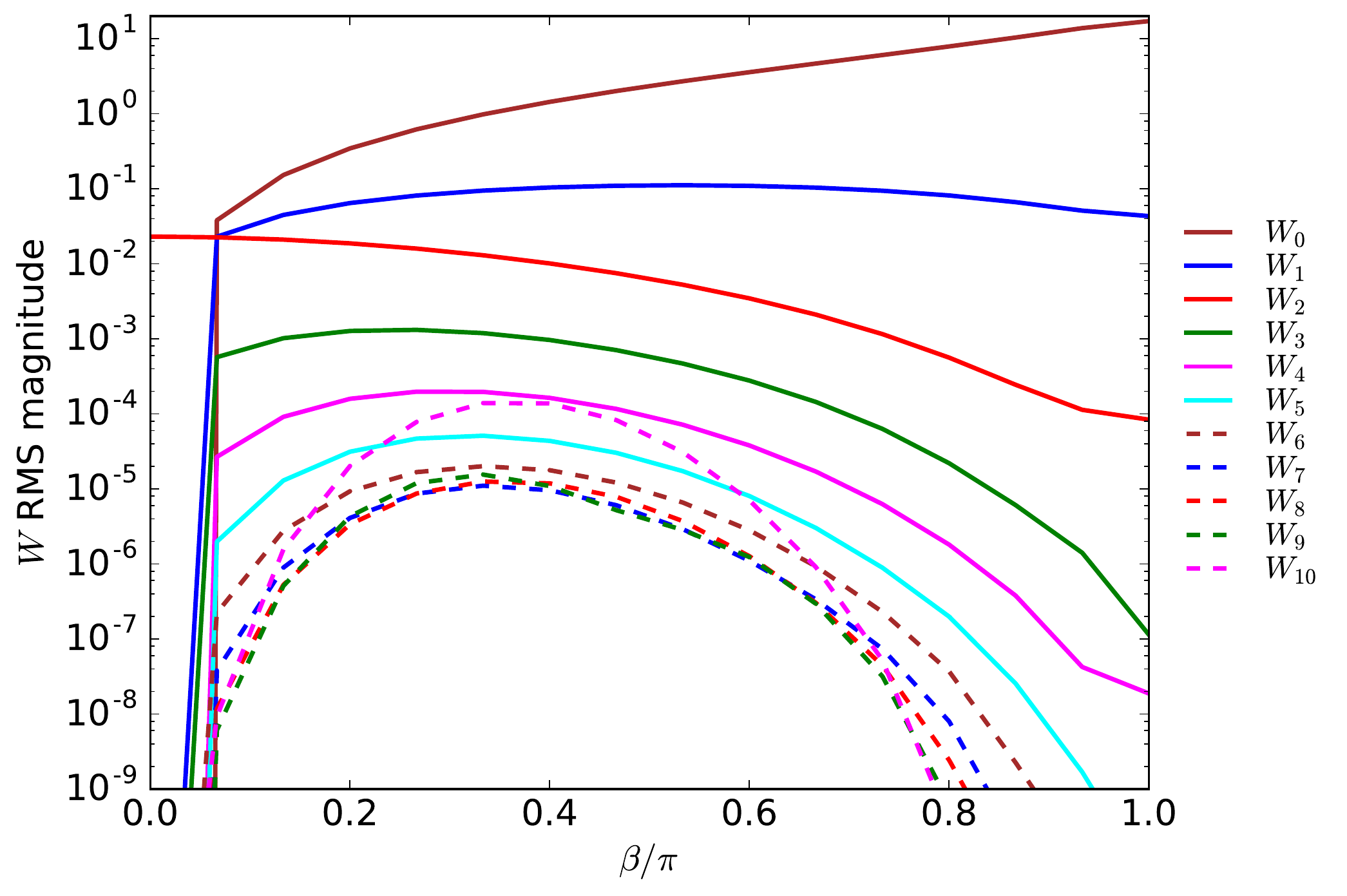}
\caption{Magnitudes of the disentangled amplitudes in the periodic half-filled 10-site Hubbard lattice.  Upper left panel: Hubbard $U/t = 2.2$. Upper right panel: $U/t = 4$. Lower left panel: $U/t = 10$. Lower right panel: $U/t = 20$. 
\label{Fig:Utilde}}
\end{figure*}

\subsubsection{Differential Equation for Disentangled Cluster Operators}
For the special case of unrestricted spin projection, we can solve for the disentangled cluster operators $W$ by solving a set of differential equations with the spin rotation angle $\beta$ as the independent variable.  By integrating this set of ordinary differential equations (ODEs), we can in principle obtain exact untruncated amplitudes, though in practice other approximations must be made.

To proceed, we need the derivative of $V_1$ with respect to $\beta$.  We define
\begin{equation}
\frac{\mathrm{d} \hfill}{\mathrm{d}\beta} V_1 = (\mathrm{e}^{V_1} \, (-\mathrm{i} \, S_y) \, \mathrm{e}^{-V_1})_d \equiv X
\label{Eqn:DefX}
\end{equation}
where the subscript $d$ means only the deexcitation part is retained and where this defines $X$.  We derive this equation in Appendix \ref{Appendix:DerivV}.

Now we note that
\begin{equation}
\mathrm{e}^{W_0} = \langle\phi| \mathrm{e}^{V_1} \, \mathrm{e}^{U_2} |\phi\rangle.
\end{equation}
Differentiating with respect to $\beta$ leads to
\begin{align}
\mathrm{e}^{W_0} \, \frac{\mathrm{d}\hfill}{\mathrm{d}\beta} W_0
 &= \langle\phi| X \, \mathrm{e}^{V_1} \, \mathrm{e}^{U_2} |\phi\rangle
\\
 &= \langle\phi| X \, \mathrm{e}^{W_0+W_1+\ldots} |\phi\rangle 
\nonumber
\\
 &= \mathrm{e}^{W_0} \, X^i_a \, W^a_i
\nonumber
\end{align}
whence
\begin{equation}
\frac{\mathrm{d} \hfill}{\mathrm{d}\beta} W_0 = X^i_a \, W^a_i.
\end{equation}
Given $W_1$, we can integrate this equation to obtain $W_0$.

\begin{figure*}[t]
\includegraphics[width=0.96\columnwidth]{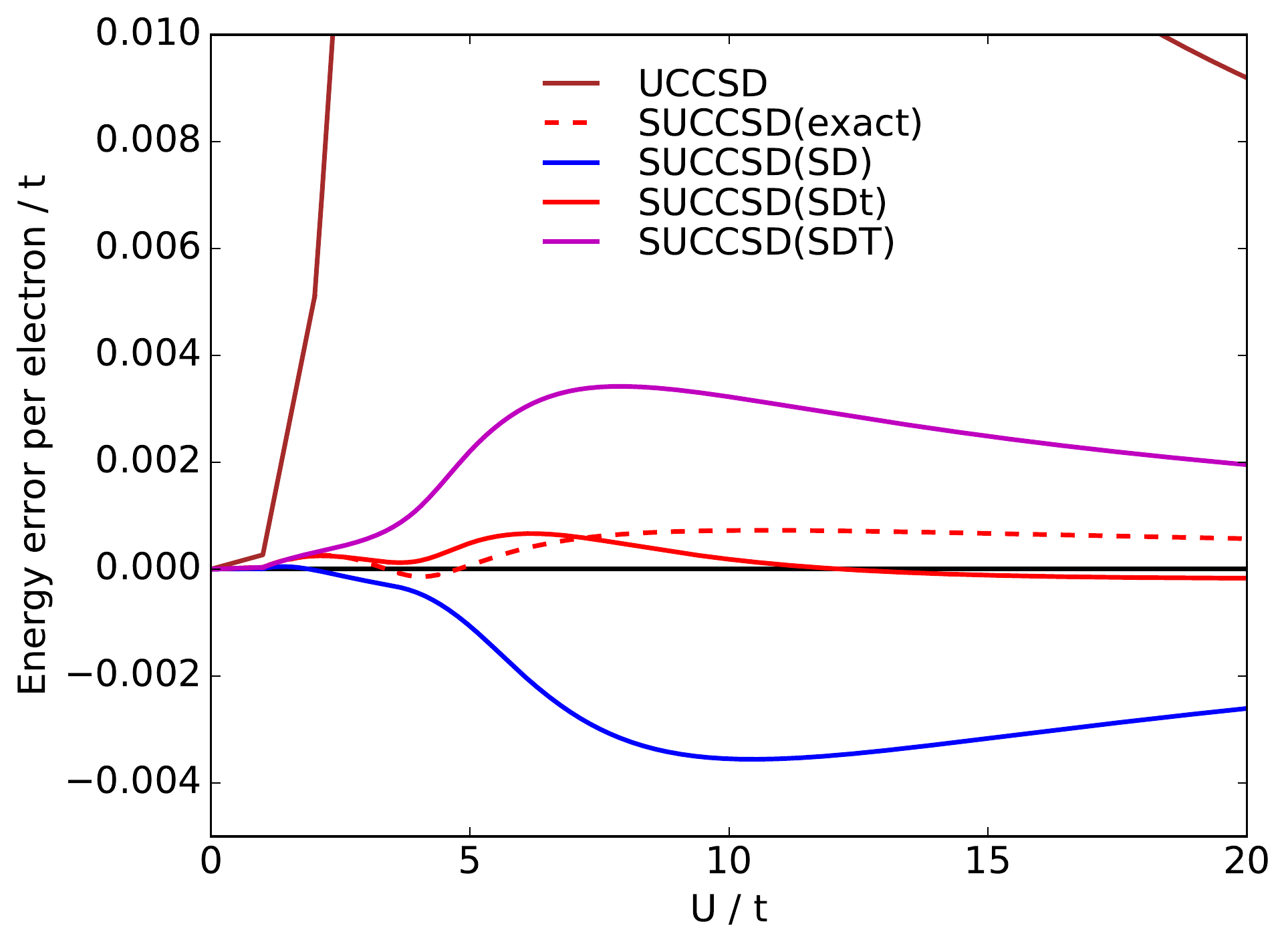}
\hfill
\includegraphics[width=0.96\columnwidth]{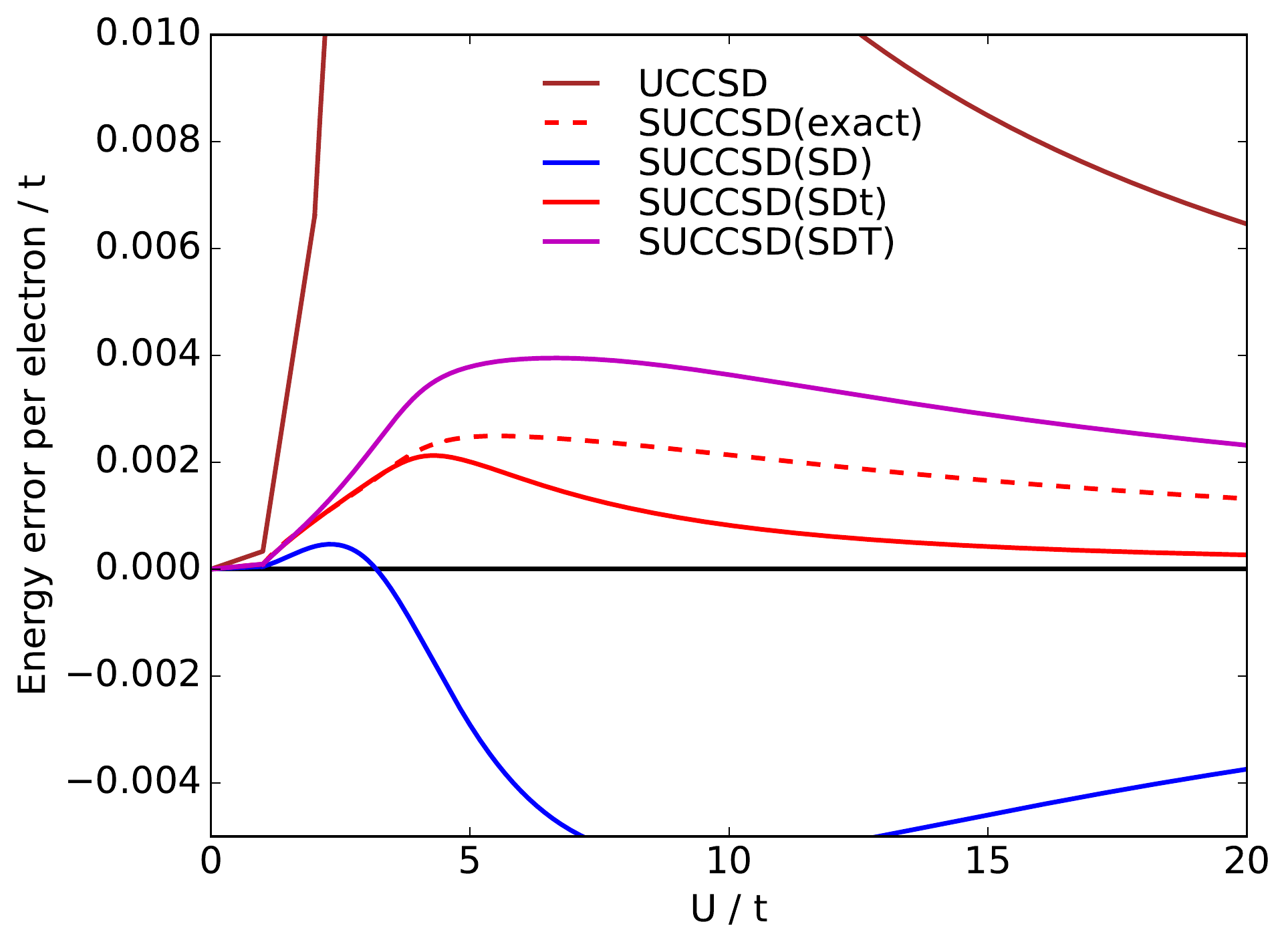}
\caption{Comparison of various truncation schemes of the PAV energy functional without response on periodic half-filled Hubbard lattices. Left panel: 6-site Hubbard model. Right panel: 10-site Hubbard model.  
\label{Fig:PbarHbarTrun}}
\end{figure*}

\begin{figure*}[t]
\includegraphics[width=0.96\columnwidth]{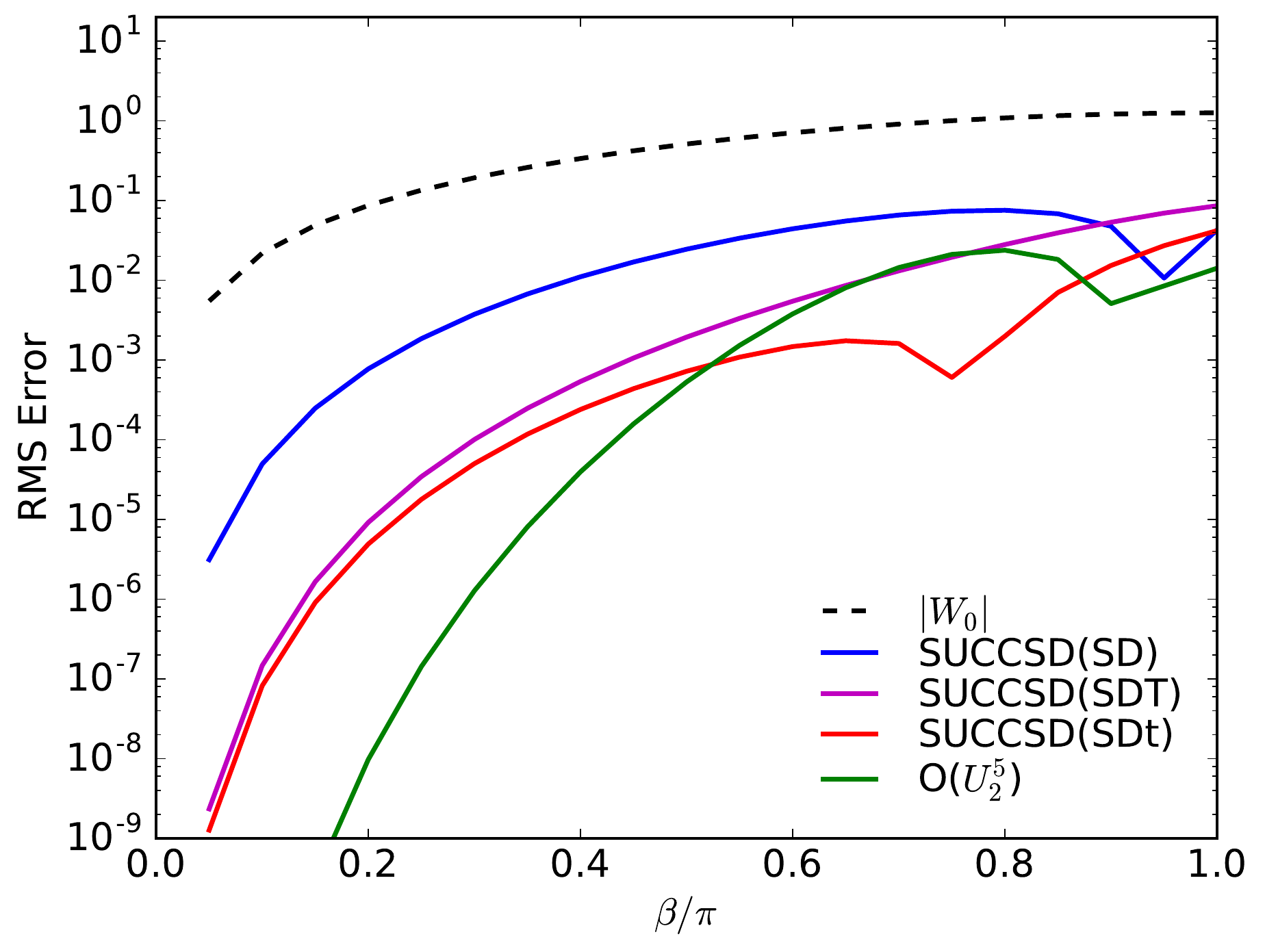}
\hfill
\includegraphics[width=0.96\columnwidth]{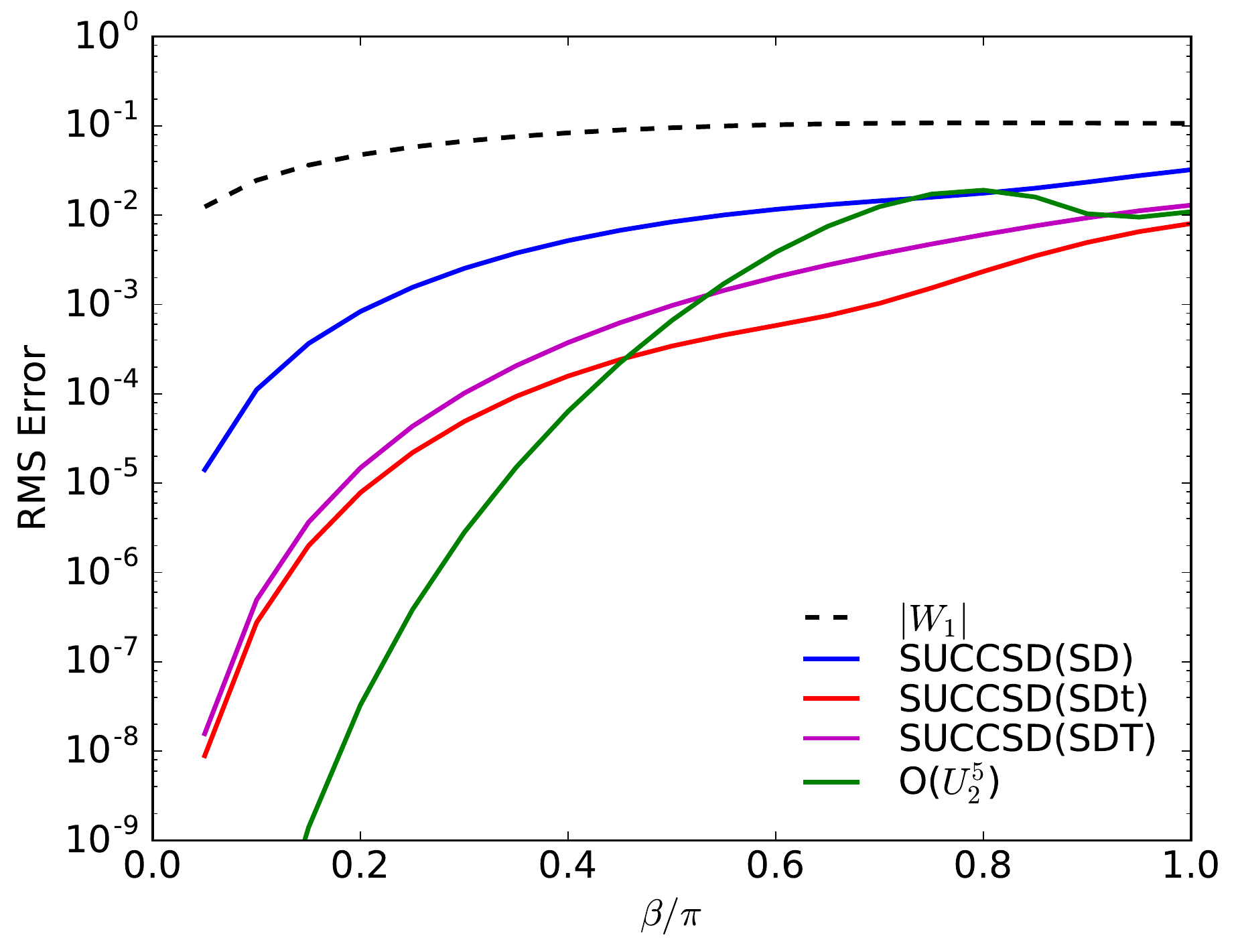}
\caption{Root-mean-square errors of disentangled amplitudes from different approximation schemes tested in the periodic half-filled 6-site Hubbard lattice with Hubbard $U/t = 4$. Left panel: Error of $W_0$.  Right panel: Error of $W_1$.   We also show the root-mean-square values of the disentangled amplitudes as $|W_0|$ and $|W_1|$.  Corresponding figures for $U/t = 20$ and errors in $W_2$ are shown in the Supplementary Material.  
\label{Fig:UtildeErr}}
\end{figure*}

Similarly, we have
\begin{equation}
\mathrm{e}^{W_0} \, W^a_i  = \langle\phi^a_i| \mathrm{e}^{V_1} \, \mathrm{e}^{U_2} |\phi\rangle.
\end{equation}
Differentiating both sides leads to
\begin{align}
\frac{\mathrm{d}\hfill}{\mathrm{d}\beta} \mathrm{e}^{W_0} \, W_i^a
 &= \mathrm{e}^{W_0} \, \frac{\mathrm{d}\hfill}{\mathrm{d}\beta} W^a_i
  + W^a_i \, \mathrm{e}^{W_0} \, \frac{\mathrm{d}\hfill}{\mathrm{d}\beta} W_0
\\
 &= \langle\phi^a_i| X \, \mathrm{e}^{V_1} \, \mathrm{e}^{U_2} |\phi\rangle
\nonumber
\\
 &= \langle\phi^a_i| X \, \mathrm{e}^{W_0+W_1+\ldots} |\phi\rangle 
\nonumber
\\
 &= \mathrm{e}^{W_0} ( W^a_i \, X^k_c \, W^c_k - W^a_k \, X^k_c \, W^c_i + X^k_c \, W^{ac}_{ik}).
\nonumber
\end{align}
Here, summation over repeated indices is implied.  Notice that the first term on the right-hand-side is
\begin{equation}
W^a_i \, X^k_c \, W^c_k = W^a_i \, \frac{\mathrm{d}\hfill}{\mathrm{d}\beta} W_0.
\end{equation}
Then the disconnected parts on both sides cancel and we have simply
\begin{equation}
\frac{\mathrm{d}\hfill}{\mathrm{d}\beta} W^a_i = X^k_c \, \left(W^{ac}_{ik} - W^a_k \, W^c_i\right).
\end{equation}

We can obtain a similar result for double excitations.  Starting from
\begin{equation}
\mathrm{e}^{W_0} \, (W^{ab}_{ij} + W^a_i \, W^b_j - W^a_j \, W^b_i) = \langle\phi^{ab}_{ij}| \mathrm{e}^{V_1} \, \mathrm{e}^{U_2} |\phi\rangle,
\end{equation}
we obtain
\begin{equation}
\frac{\mathrm{d\hfill}}{\mathrm{d}\beta} W^{ab}_{ij} = X^k_c \, \left(W_{ijk}^{abc} - W^{ab}_{kj} \, W^c_i - W_{ij}^{cb} \, W^a_k\right).
\end{equation}

Let us note a few important features of these differential equations for the disentangled cluster operators.  First, at $\beta = 0$ we have $V_1 = 0$, so $W_0 = 0$, $W_1 = 0$, and $W_2 = U_2$.  Second, the derivative of $W_k$ contains $W_{k+1}$ but no higher excitations.  We can thus approximate $W_{k+1}$ for some excitation level $k$ and from there solve for the lower excitation levels.  For this reason we prefer the differential equations to the algebraic equations given in the previous section, as the latter require both $W_{k+1}$ and $W_{k+2}$ to obtain $W_k$; morever, the algebraic equations do not allow us to compute $W_0$.

Here, we consider several approximations.  If we make the approximation $W_3 = 0$ and integrate to obtain $W_0$ to $W_2$, we call the resulting method SUCCSD(SD) where SUCCSD stands for spin-projected unrestricted CCSD and ``(SD)'' emphasizes that we retain through $W_2$ in the differential equation.  Similarly, SUCCSD(SDT) means we set $W_4 = 0$ and solve for $W_0$ to $W_3$.   The cost per grid point of SUCCSD(SD) is $\mathcal{O}(N^6)$, and that of SUCCSD(SDT) is $\mathcal{O}(N^7)$.  A third approach we consider is to solve for an approximate $W_3$ from the algebraic equations, by setting $W_3$ through $W_5$ to zero in the right-hand-side of Eqn. \ref{Eqn:Alg2}, thereby obtaining an approximation for $W_3$ in terms of $W_1$ and $W_2$.  We call the resulting method SUCCSD(SDt), and it scales as $\mathcal{O}(N^6)$.  Because including linear response in the energy expression or solving the amplitude equations would require through $W_4$, in this work we use the ODE technique only to evaluate the energy in a PAV approach without response.  This allows us to focus on assessing the error introduced by our approximations for $W_{k+1}$.

\subsection{Results}
The key objects of practical implementations of our spin projected coupled cluster theory are the disentangled cluster operators $W_k$.  All our approximations are based on assuming that $W_k$ for sufficiently high excitation level are negligible or at least readily approximated by the $W_k$ for lower excitation levels.  It is thus critical to check that this is indeed the case.  Figure \ref{Fig:Utilde} shows the behavior of $W_k$ for the ten-site half-filled periodic Hubbard lattice at several values of interaction strength $U/t$.  As we have discussed, only $W_2$ is non-zero at $\beta = 0$.  As $\beta$ is increased, other amplitudes appear.  In general, $W_k$ indeed decays as excitation level increases for all $\beta$ and for all interaction strengths.  This justifies our various truncation schemes.  Note, however, that $W_0$ actually increases as $\beta$ increases.  We will discuss this increase later.

A different way of justifying our truncation schemes is by comparison with the untruncated results. This is done in Fig. \ref{Fig:PbarHbarTrun}.  Unsurprisingly, SUCCSD(SDT) is closer to the exact SUCCSD than is SUCCSD(SD).  More interestingly, SUCCSD(SDt) is closer to the exact result than is SUCCSD(SDT).  While this may seem counterintuitive, it ultimately arises from the observation that the algebraic equations lead to a more accurate approximation for $W_3$ than does solving the differential equation assuming $W_4 = 0$.  We should note that both SUCCSD(SD) and SUCCSD(SDt) scale the same as UCCSD, and both are significantly more accurate.  

To bear out our explanation for the superiority of SUCCSD(SDt) over SUCCSD(SDT), Fig. \ref{Fig:UtildeErr} shows that SUCCSD(SDt) produces more accurate disentangled amplitudes (as a consequence of which it also produces more accurate kernels, as shown in the Supplementary Material).  Figure \ref{Fig:UtildeErr} also shows the behavior of the order-by-order expansion of the disentangled amplitudes.  At small $\beta$ where $V_1$ is small, this expansion works exceptionally well, but as $\beta$ increases its quality deteriorates rapidly.

\begin{figure}
\includegraphics[width=0.96\columnwidth]{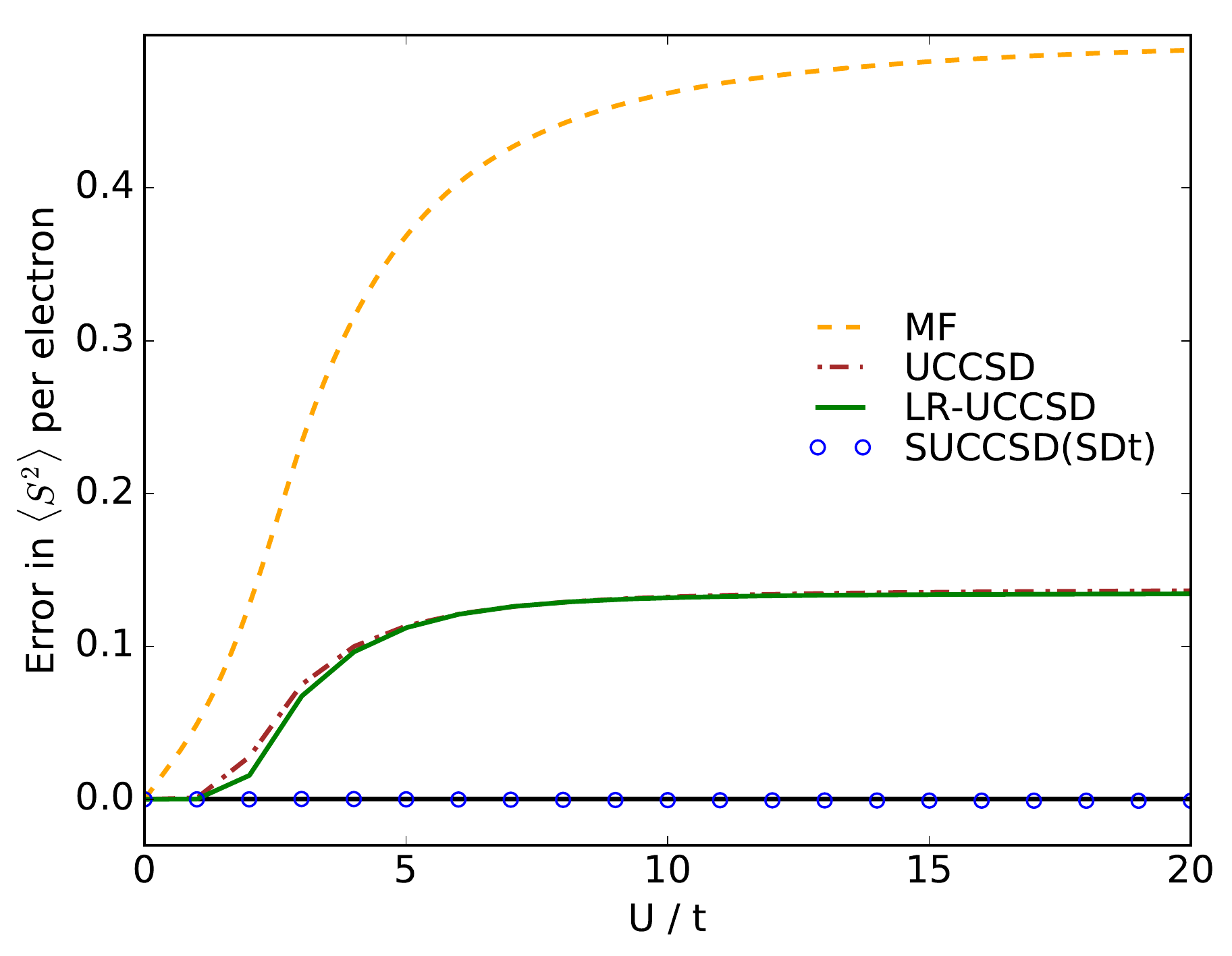}
\caption{Error per electron in the expectation value of $S^2$ for the periodic half-filled 18-site Hubbard lattice.  We use the broken-symmetry determinant of SUHF as a reference so that the mean-field and the UCCSD break spin symmetry everywhere.  The error in the expectation value from the mean-field reference is ``MF.''
\label{Fig:Spin}}
\end{figure}

One might be concerned that approximating the disentangled cluster operators spoils the symmetry projection.  After all, approximating $W$ implies approximately handling $V_1$, hence approximately handly the rotation operator $R$ and hence approximating the symmetry projection.  Figure \ref{Fig:Spin} shows that, at least with our SUCCSD(SDt) approximation, this is not a concern.  Here, we plot the error in the expectation value of $\langle S^2\rangle$ per electron in the periodic half-filled eighteen-site Hubbard lattice.  We use the reference determinant of SUHF as our mean-field reference so the mean-field breaks symmetry everywhere.  

To be clear about what we plot, the correct definition of the coupled cluster expectation value of an operator $\mathcal{O}$ is
\begin{equation}
\langle \mathcal{O} \rangle_\mathrm{CC} = \langle \phi| (1 + Z) \, \mathrm{e}^{-U} \, \mathcal{O} \, \mathrm{e}^U |\phi\rangle.
\end{equation}
This is what we have plotted as ``LR-UCCSD'' where $\mathcal{O} = S^2$.  Neglecting $Z$ gives us the curve we have plotted as ``UCCSD.''  For projected coupled cluster, the expectation value would similarly be
\begin{equation}
\langle \mathcal{O} \rangle_\mathrm{PCC} = \frac{\langle \phi| (1 + Z) \, \mathrm{e}^{-U} \, P \, \mathcal{O} \, \mathrm{e}^U |\phi\rangle}{\langle \phi| (1+Z) \, \mathrm{e}^{-U} \, P \, \mathrm{e}^U |\phi\rangle}.
\end{equation}
Evaluating this projected coupled cluster expectation value in the presence of $Z$ would require excited kernels which we have not yet constructed, so we have plotted it without $Z$, as ``SUCCSD(SDt).''  If the projection is done exactly, the expectation value of $S^2$ would be exact even in the absence of $Z$.

While the mean-field is badly symmetry broken for large Hubbard parameter $U/t$, unrestricted CCSD greatly reduces the degree of symmetry breaking.  Interestingly, including linear response has minimal effect in this case.  Our SUCCSD(SDt) gives almost exact symmetry projection, with error per electron in the expectation value of $S^2$ on the order of $10^{-3}$ (and very slightly negative in places, which is possible because we do not use a Hermitian expectation value).

\subsection{The Thermodynamic Limit}
\begin{figure*}[t]
\includegraphics[width=0.96\columnwidth]{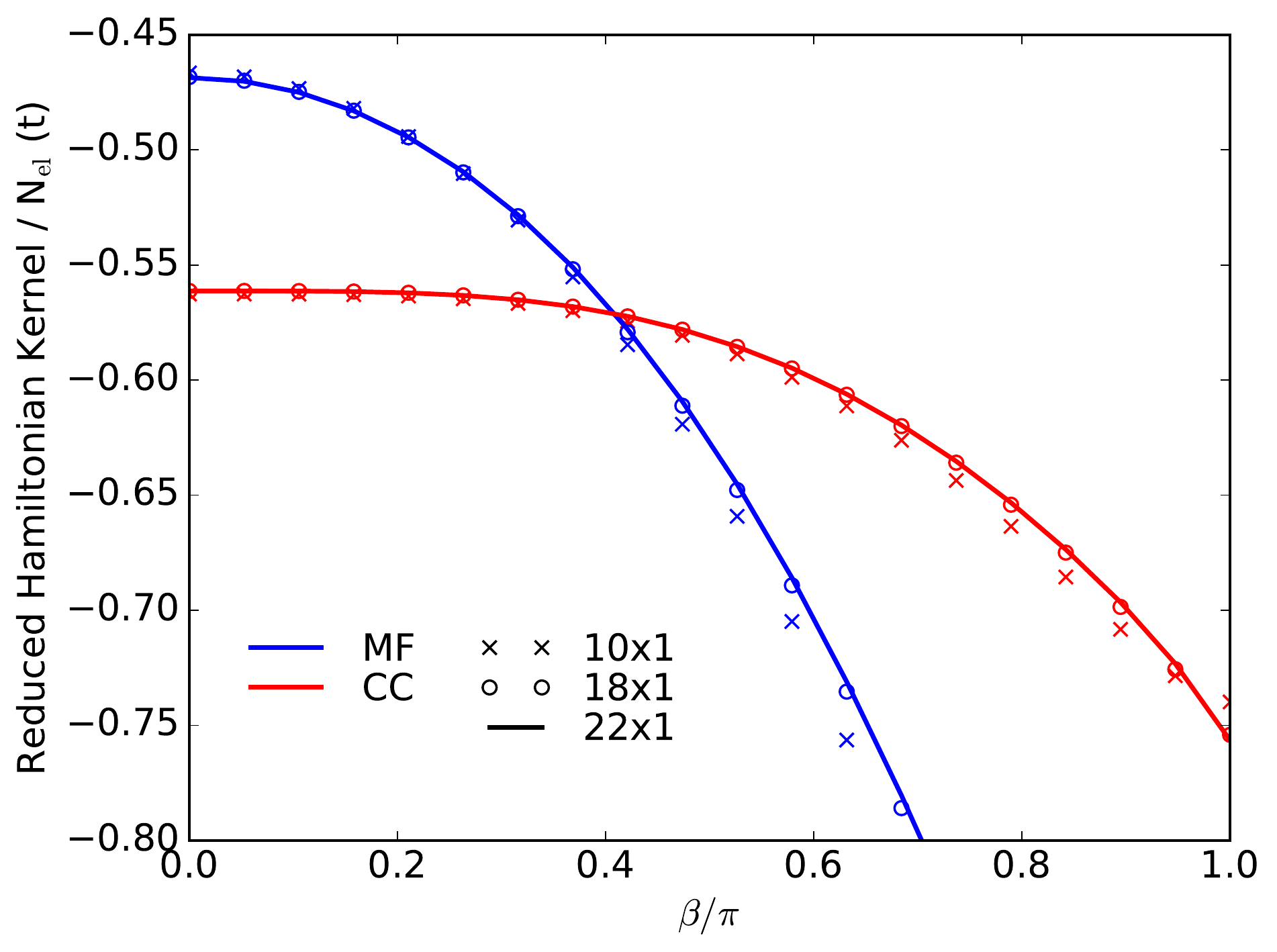}
\hfill
\includegraphics[width=0.96\columnwidth]{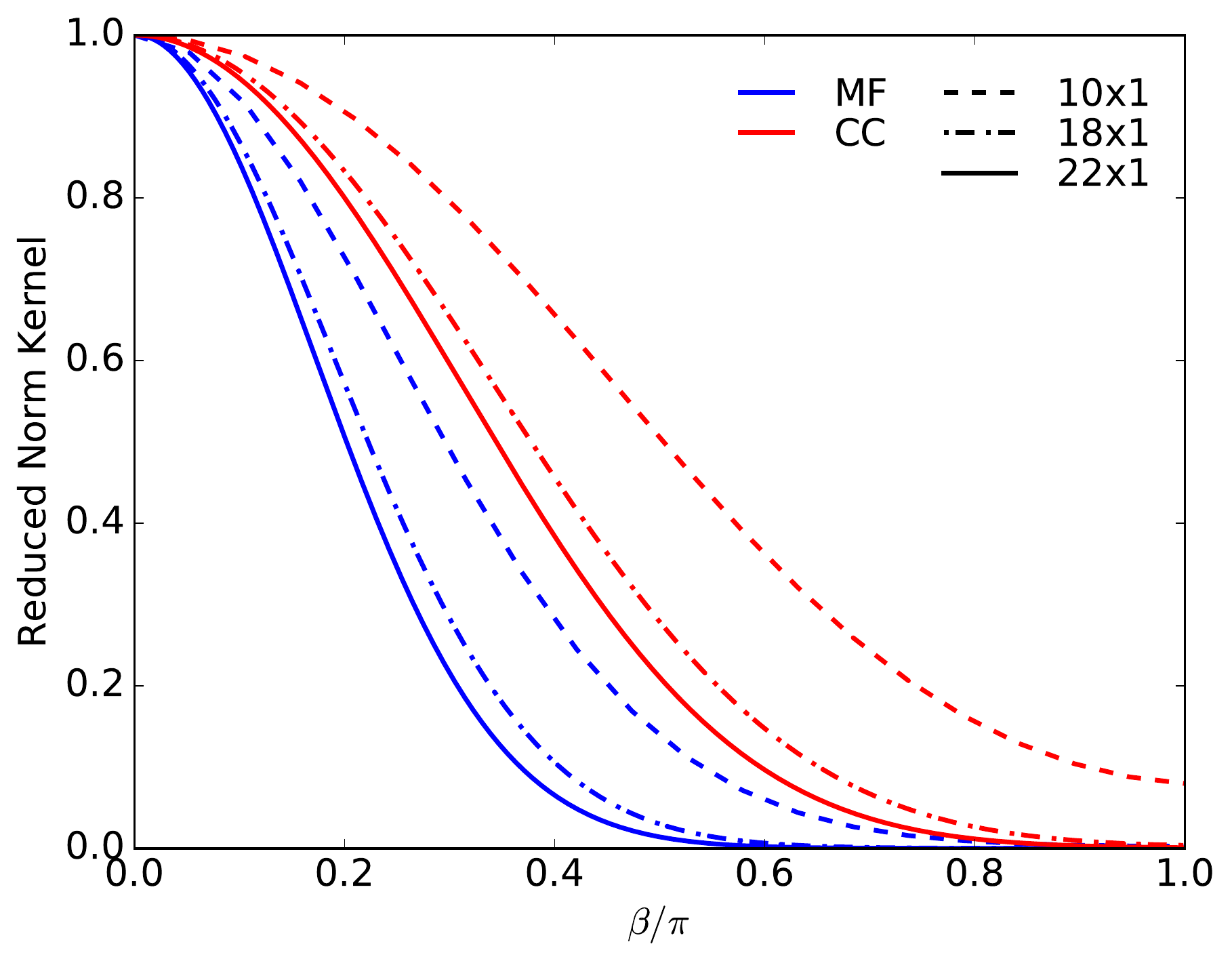} 
\caption{Norm and Hamiltonian kernels for the one-dimensional periodic half-filled Hubbard lattice at different system sizes. Left panel: Hamiltonian kernel at Hubbard $U/t$ = 4. Right panel: Norm kernel at $U/t$ = 4. Corresponding figures for $U/t = 20$ are shown in the Supplementary Material.
\label{Fig:KernelDiffSize}}
\end{figure*}

\begin{figure}[t]
\includegraphics[width=\columnwidth]{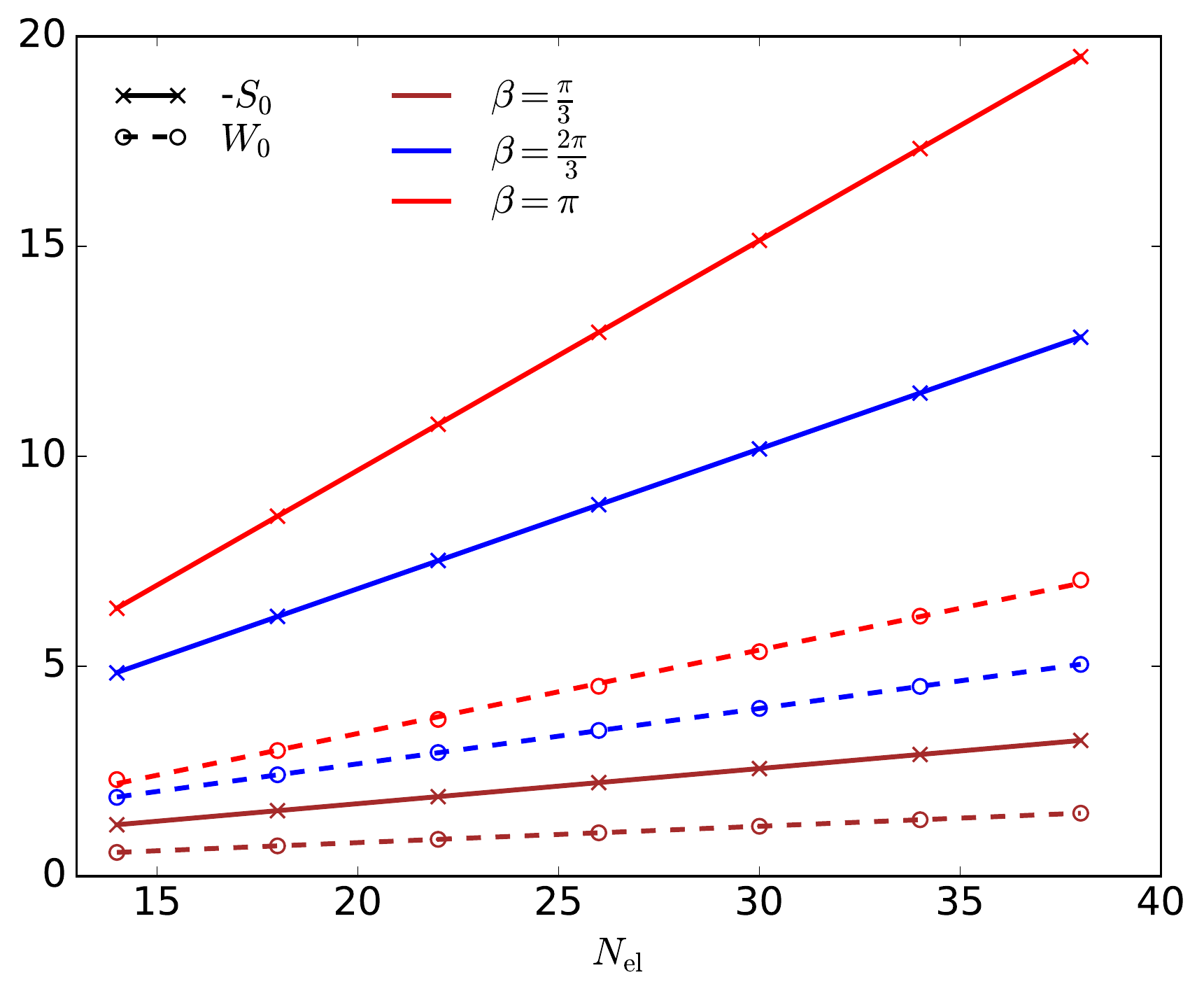} 
\caption{Scaling of $W_0$ and the mean-field norm kernel exponent $S_0$ with respect to system size at Hubbard $U/t = 4$. The corresponding figure for $U/t = 20$ is shown in the Supplementary Material. 
\label{Fig:Scale}}
\end{figure}

\begin{figure*}[t]
\includegraphics[width=0.96\columnwidth]{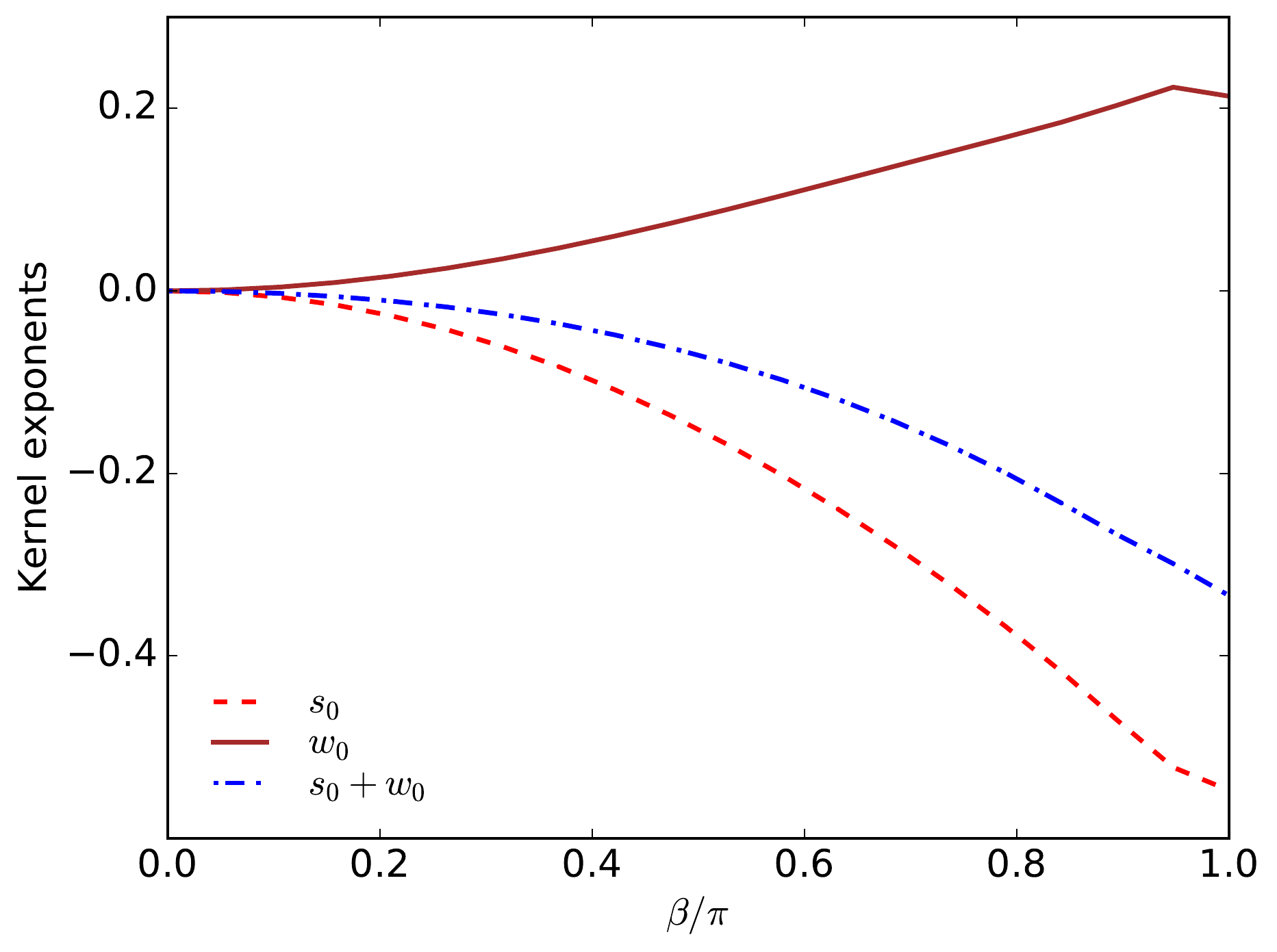}
\hfill
\includegraphics[width=0.96\columnwidth]{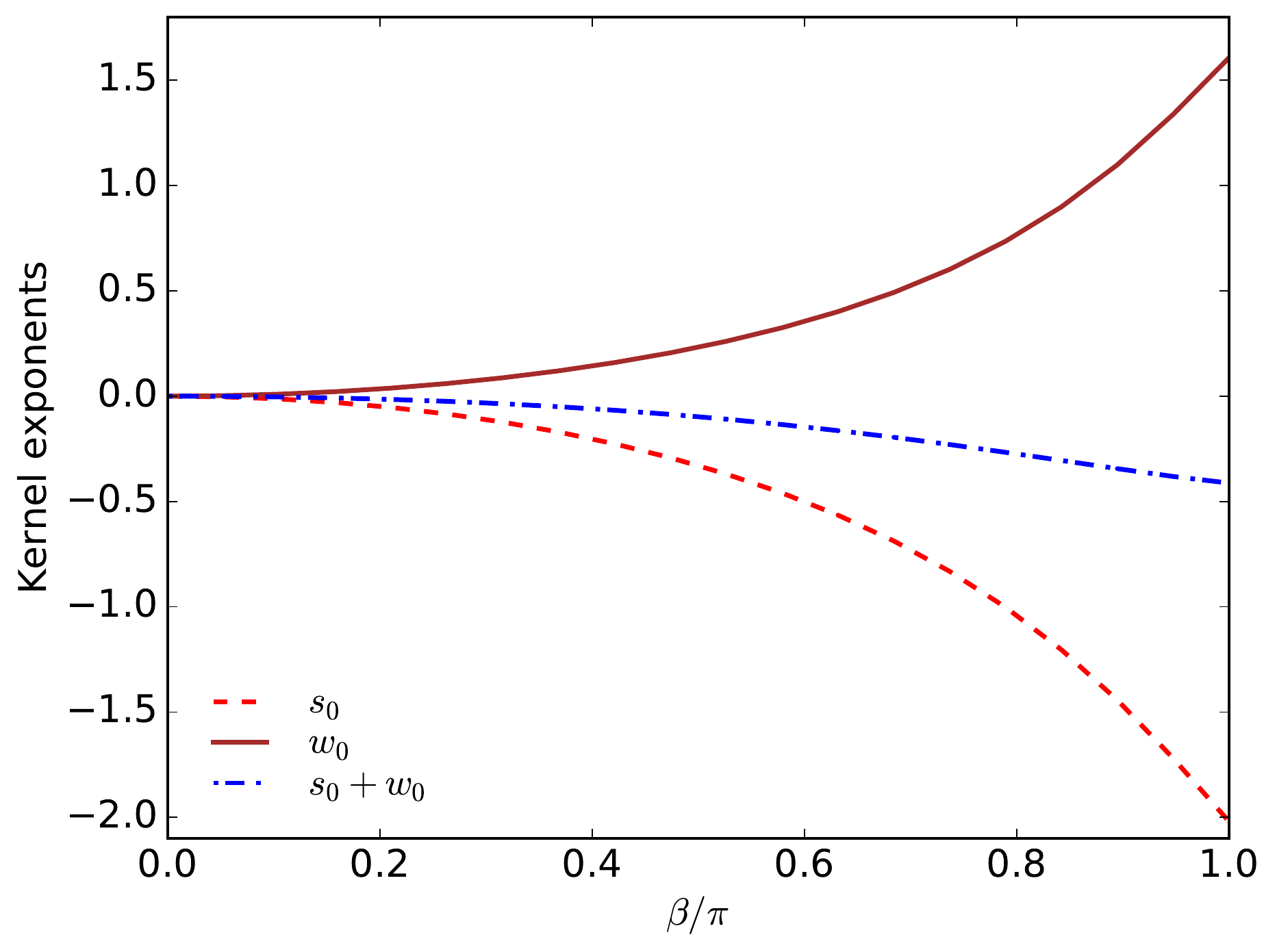}
\caption{Scaling factor of the mean-field kernel and $W_0$ as a function of $\beta$ for one-dimensional periodic half-filled Hubbard lattices. 
Left panel: Hubbard $U/t = 4$.  Right panel: $U/t = 20$.  Corresponding figures for $U/t = 8$ and $U/t = 12$ are shown in the Supplementary Material.
\label{Fig:ScaleFac}}
\end{figure*}

\begin{figure}[t]
\includegraphics[width=\columnwidth]{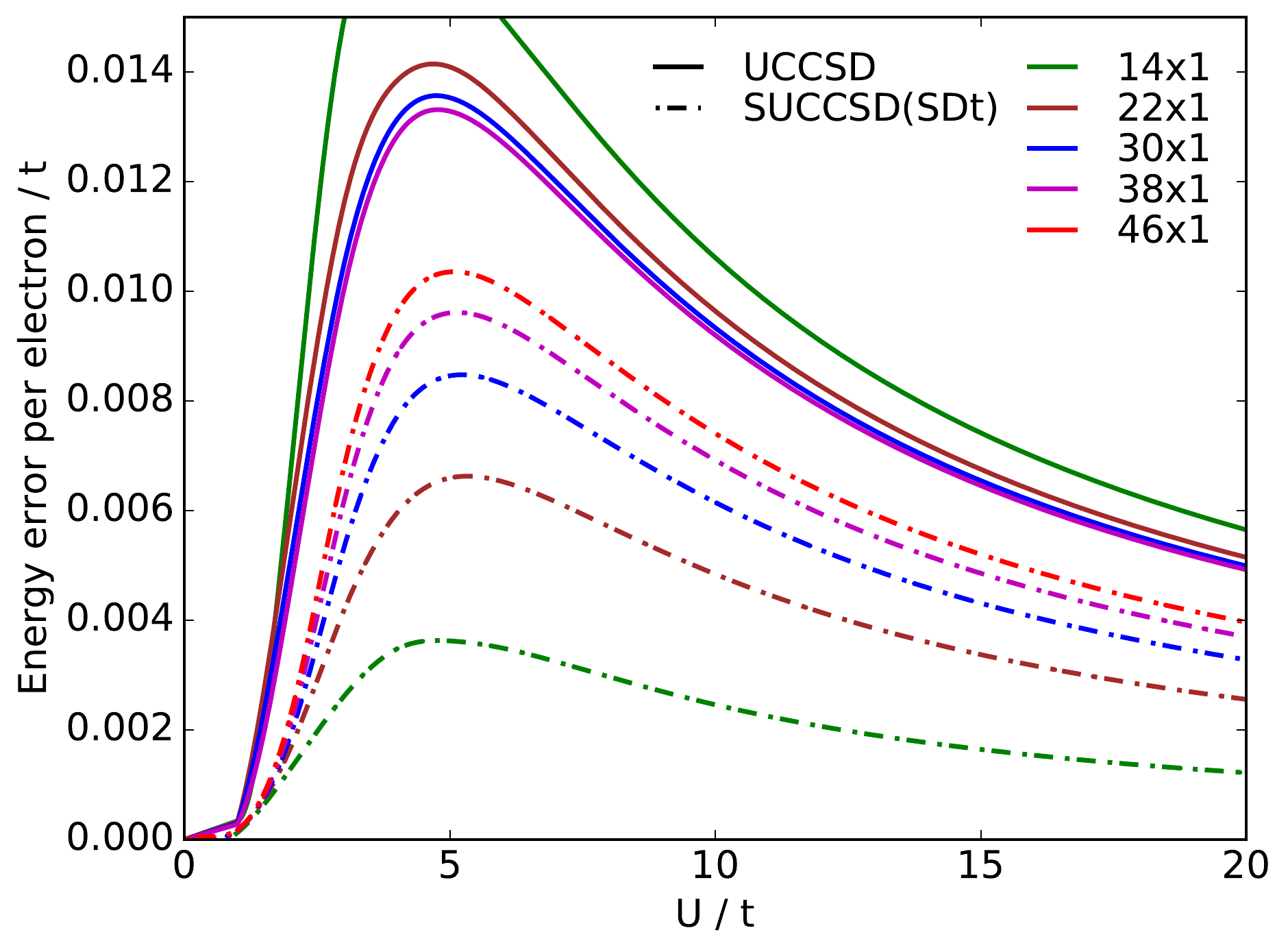}
\caption{PAV energy errors per electron with respect to the exact result in half-filled one-dimensional Hubbard rings. 
\label{Fig:HubDiffSize}}
\end{figure}

It is known that spin projected UHF is not size-extensive, or, more precisely, that its size-extensive energetic component is the same as that of the broken symmetry mean-field.\cite{PHF}  Here, we demonstrate numerically that SUCCSD likewise has no size extensive correction beyond UCCSD, but that the SUCCSD energy returns to the UCCSD value at a much slower rate than the SUHF energy returns to UHF.  This bodes well for large finite systems, and we expect SUCCSD to provide useful improvements to UCCSD even when SUHF provides only marginal improvements upon UHF.  We should note that our results here are generated for the SUCCSD(SDt) approximation to the exact SUCCSD, and in the PAV sense, but we expect the exact VAP SUCCSD to display broadly similar features although we have not proven as much.

From the differential equation one can see that the derivatives of the $W_k$ contain only connected terms.  At $\beta=0$ where $V_1 = 0$, $W = U_2$ and is extensive.  This suggests that $W$ at every $\beta$ should be extensive and $W$ should converge with respect to system size at a rate similar to UCCSD.  Since the reduced Hamiltonian kernel contains only connected terms involving $W$, it, too, should be size extensive.  This is indeed the case, as can be seen from Fig. \ref{Fig:KernelDiffSize}.  The Hamiltonian kernel per electron is already converged with respect to system size for the 22-site Hubbard ring for all interaction strengths, which is where UCCSD also converges.  The SUHF Hamiltonian kernel also converges rapidly with respect to system size.  

The problem, then, must be the norm kernel, which as we see converges very slowly.  In fact, in the thermodynamic limit the norm kernel for SUHF becomes a delta function at $\beta = 0$; thus the energy only samples $\beta = 0$, and we therefore obtain the UHF result.

In SUCC theory, the norm kernel is $\langle \phi|R(\beta)|\phi\rangle \, \mathrm{e}^{W_0}$, where $\langle \phi|R(\beta)|\phi\rangle \equiv \mathrm{e}^{S_0}$ is the SUHF norm kernel.  Figure \ref{Fig:Scale} shows how $W_0$ and $S_0$ behave as a function of system size.  It is clear that
\begin{subequations}
\begin{align}
W_0 &\approx w_0 \, N,
\\
S_0 &\approx s_0 \, N,
\end{align}
\end{subequations}
where $w_0$ and $s_0$ are constants, so that the coupled cluster norm kernel is
\begin{equation}
\langle \phi|R|\phi\rangle \, \mathrm{e}^{W_0} \approx \mathrm{e}^{(s_0 + w_0) \, N}.
\end{equation}
However, $s_0$ and $w_0$ have opposite signs; as the number of electrons increases, the SUHF norm kernel goes to zero while $W_0$ goes to infinity.  This makes it difficult to tell how the SUCC norm kernel behaves, although clearly it decays less rapidly than does the SUHF norm kernel.  To help resolve this behavior, Fig. \ref{Fig:ScaleFac} shows that $s_0$ is negative for all $\beta$, while $w_0$ is positive and thus counteracts the decay of the mean-field kernel, but their sum is always negative except at $\beta = 0$.  This means that the norm kernel in the thermodynamic limit vanishes except at $\beta = 0$, just as it does for SUHF.  Thus, SUCCSD approaches UCCSD in the thermodynamic limit, but at a much slower rate than SUHF approaches UHF (because $s_0 + w_0$ is less negative than $s_0$ alone).  Figure \ref{Fig:HubDiffSize} shows that SUCCSD adds considerable correlation atop of UCCSD even at large system sizes (see results for the 46-site lattice).

We close with a brief discussion of how the kernels should behave as the theory becomes even more complete.  In the full coupled cluster limit, the wave function, though written in the language of symmetry broken Hartree-Fock, has good quantum numbers (for finite systems).  Accordingly, the norm kernel is one for all $\Omega$ and the Hamiltonian kernel is the exact ground state energy.  Our results here show that SUCCSD has flatter norm and Hamiltonian kernels than does SUHF, and as we continue to increase the level of correlation in the cluster operator, we would expect to see the kernels become even flatter.

\section{Discussion}
\subsection{Comparison With Duguet's Symmetry Broken and Restored Coupled Cluster Theory}
Recently, Duguet proposed a symmetry broken and restored coupled cluster theory.\cite{Duguet2015}  Here, we briefly discuss the differences between our two approaches.

The biggest difference is that we begin with an explicit wave function ansatz rather than with the kernels.  As a result, the linear response and optimization of the cluster operator $U$ force us to introduce excited Hamiltonian and norm kernels which do not appear in Duguet's theory.  Second,  Duguet's approach includes solving a coupled cluster-like equation at each $\Omega$ where we solve for the cluster operator $U$ only once and use these amplitudes to generate the $\Omega$-dependent kernels.  Finally, we introduce the disentangled cluster operators $W$, which conveniently encapsulate all information from the various kernels we require.  We may solve a set of coupled differential equations for the disentangled cluster operators.  While Duguet also proposes a differential equation, it is used only for the norm kernel.

\subsection{Conclusions}
This work combines coupled cluster and symmetry projection into one tool which seems to have considerable promise.  The broken symmetry coupled cluster is already fairly accurate for many cases, and the symmetry restored coupled cluster is even better while retaining good symmetries.  Even the simple projection after variation scheme is a useful improvement on broken symmetry coupled cluster, and results can be improved further by including linear response or by reoptimizing the broken symmetry cluster operator in the presence of the symmetry projector.  For the small half-filled Hubbard rings for which we have been able to carry out the exact VAP SUCCSD calculations, errors are well under $0.001 t$ per electron.

Symmetry projected coupled cluster reduces energetically to broken symmetry coupled cluster in the thermodynamic limit, just as projected Hartree-Fock reduces to the broken symmetry mean field.  However, the symmetry projected coupled cluster does so more slowly, and useful improvements upon broken symmetry coupled cluster are possible even for fairly sizable systems.

An important quantity introduced in this work is the disentangled cluster operators $W$.  These allow us to work in the same orthogonal excitation framework as traditional coupled cluster theory.  The needed kernels can be evaluated from disentangled cluster operators of low excitation rank.  The disentangled cluster operators decay as the excitation level increases and, for unrestricted spin projection, satisfy a set of simple differential equations which couple $W_k$ of different excitation levels.  We have shown several truncation schemes for the approximate construction of the disentangled cluster operators.

Much work, of course, remains.  We have by no means explored all possible ways of approximating the disentangled cluster operators, nor have we used them in the construction of excited kernels, which one must do if one wishes to make linear response and VAP calculations practical.  Neither have we implemented the combination of spin projected coupled cluster based on a generalized Hartree-Fock reference which breaks both $S^2$ and $S_z$ symmetries, which we expect it to be significantly more accurate yet.\cite{SGCCSD}  Thus far we have considered only the ground state energy, and naturally properties, gradients, and excited states can all in principle be accessed by suitable modifications of traditional coupled cluster methods.  Nonetheless, we are greatly encouraged by this early foray into the symmetry projection of broken symmetry coupled cluster theory.

\section*{Supplementary Material}
See supplementary material for data corresponding to Figs. \ref{Fig:UtildeErr}, \ref{Fig:KernelDiffSize}, \ref{Fig:Scale}, and \ref{Fig:ScaleFac} at other values of Hubbard parameters $U/t$.  We also show errors in the disentangled cluster operator $W_2$ and the norm and Hamiltonian kernels corresponding to different approximations to SUCCSD.

\begin{acknowledgments}
This work was supported by the National Science Foundation under award CHE-1462434. G.E.S. is a Welch Foundation Chair (C-0036).  We would like to acknowledge computational support provided by the Center for the Computational Design of Functional Layered Materials, an Energy Frontier Research Center funded by the U.S. Department of Energy, Office of Science, Basic Energy Sciences under Award DE-SC0012575.
\end{acknowledgments}

\appendix
\section{Non-Zero Components of Auxilliary Kernels \label{Appendix:NonZeroNH}}
In Eqn. \ref{Eqn:AuxilliaryKernels} we introduced the auxiliary kernels $\tilde{\mathcal{N}}$ and $\tilde{\mathcal{H}}$.  Here we list their nonzero elements for convenience.  They are
\begin{subequations}
\begin{align}
\tilde{\mathcal{N}}_i^i &= 1,
\\
\tilde{\mathcal{N}}_i^a &= \frac{\langle \phi_i^a| \mathrm{e}^{V_1} \, \mathrm{e}^U |\phi\rangle}{\mathcal{N}},
\\
\tilde{\mathcal{N}}_{ij}^{ij} &= -\tilde{\mathcal{N}}_{ij}^{ji} = 1 - \delta_{ij},
\\
\tilde{\mathcal{N}}_{ij}^{aj} &= \tilde{\mathcal{N}}_{ji}^{ja} = -\tilde{\mathcal{N}}_{ij}^{ja} = -\tilde{\mathcal{N}}_{ji}^{aj}  =\left(1 - \delta_{ij}\right) \, \tilde{\mathcal{N}}_i^a,
\\
\tilde{\mathcal{N}}_{ij}^{ab} &=  \frac{\langle \phi_{ij}^{ab}| \mathrm{e}^{V_1} \, \mathrm{e}^U |\phi\rangle}{\mathcal{N}},
\\
\tilde{\mathcal{H}}_i^i &= \mathcal{H},
\\
\tilde{\mathcal{H}}_i^a &= \frac{\langle \phi_i^a| \bar{H}_{V_1} \, \mathrm{e}^{V_1} \, \mathrm{e}^U |\phi\rangle}{\mathcal{N}},
\\
\tilde{\mathcal{H}}_{ij}^{ij} &= -\tilde{\mathcal{H}}_{ij}^{ji} = \mathcal{H} \, \left(1 - \delta_{ij}\right),
\\
\tilde{\mathcal{H}}_{ij}^{aj} &= \tilde{\mathcal{H}}_{ji}^{ja} = -\tilde{\mathcal{H}}_{ij}^{ja} = -\tilde{\mathcal{H}}_{ji}^{aj} =\left(1 - \delta_{ij}\right) \, \tilde{\mathcal{H}}_i^a,
\\
\tilde{\mathcal{H}}_{ij}^{ab} &=  \frac{\langle \phi_{ij}^{ab}| \bar{H}_{V_1} \, \mathrm{e}^{V_1} \, \mathrm{e}^U |\phi\rangle}{\mathcal{N}}.
\end{align}
\end{subequations}

\section{Including $U_1$ in the Disentangled Cluster Formalism \label{Appendix:U1}}
In our disentangled cluster formalism we incorporate $U_1$ by using it to transform the Hamiltonian and symmetry projectors rather than explicitly including it in our equations.  Here, we show how we do this.  We will use the singly-excited Hamiltonian kernel $\mathcal{H}_i^a$ to demonstrate the general idea.

The excited Hamiltonian kernel we need is
\begin{subequations}
\begin{align}
\mathcal{N} \, \mathcal{H}_i^a
 &= \langle \phi_i^a |R \, H \, \mathrm{e}^{U_1} \, \mathrm{e}^{U_2} |\phi\rangle
\\
 &= \langle \phi| c_i^\dagger \, c_a \, R \, H \, \mathrm{e}^{U_1} \, \mathrm{e}^{U_2} |\phi\rangle
\\
 &= \langle \phi| \mathrm{e}^{-U_1} \, c_i^\dagger \, c_a \, \mathrm{e}^{U_1} \, \bar{R}_{U_1} \, \bar{H}_{U_1} \, \mathrm{e}^{U_2} |\phi\rangle
\\
 &= (\tau^{-1})^a_p \, \langle \phi| c_q^\dagger \, c_p \, \bar{R}_{U_1} \, \bar{H}_{U_1} \, \mathrm{e}^{U_2}|\phi\rangle \, \tau^q_i
\end{align}
\end{subequations}
where $\bar{R}_{U_1}$ and $\bar{H}_{U_1}$ are
\begin{subequations}
\begin{align}
\bar{R}_{U_1} &= \mathrm{e}^{-U_1} \, R \, \mathrm{e}^{U_1},
\\
\bar{H}_{U_1} &= \mathrm{e}^{-U_1} \, H \, \mathrm{e}^{U_1}.
\end{align}
\end{subequations}
The orbital transformation matrix $\bm{\tau}$ is
\begin{equation}
\bm{\tau} = \begin{pmatrix} \mathbf{1} & \mathbf{0} \\ -\mathbf{U}_\mathrm{vo} & \mathbf{1} \end{pmatrix}
\end{equation}
where $\mathbf{U}_\mathrm{vo}$ is the matrix of $U_1$ amplitudes.  The practical result is that the integrals defining the Hamiltonian and rotation operator should be replaced by the transformed ones and, this having been done, the transformation matrix $\mathbf{T}$ of Eqn. \ref{Eqn:DefTMat} should be replaced by $\mathbf{T} \, \bm{\tau}$.

\section{Proof of the Disentanglement Equation \label{Appendix:ProveW}}
Here we prove our formula for the disentangled cluster operators given in Eqn. \ref{Eqn:WDef2}.

We begin by writing the wave function $|\psi\rangle = \mathrm{e}^{V_1} \, \mathrm{e}^{U_2} |\phi\rangle$ in two different ways:
\begin{subequations}
\begin{align}
\mathrm{e}^{V_1} \, \mathrm{e}^{U_2} |\phi\rangle
 &= \mathrm{e}^{W_0 + W_1 + \ldots} |\phi\rangle
\\
 &= \mathrm{e}^{W_0} \, \left(1 + C_1 + C_2 + \ldots \right) |\phi\rangle.
\end{align}
\end{subequations}
The excitation operators $C_k$ are given by the linked excitation components of
\begin{equation}
\sum_n \frac{1}{(2n-k)!} \frac{1}{n!} V_1^{2n-k} U_2^n |\phi\rangle
\end{equation}
where by ``excitation components'' we mean the maximally contracted part (so that all indices on the $V_1$ are contracted with indices on $U_2$).  The cluster operator $W_k$ is the fully connected part of $C_k$, which proves Eqn. \ref{Eqn:WDef2} for $k > 0$.  The case $k=0$ needed for $W_0$ is a special case of the linked diagram theorem.

\section{Proof of the Derivative of $V_1$ \label{Appendix:DerivV}}
In Eqn. \ref{Eqn:DefX} we write the derivative of the operator $V_1$ with respect to $\beta$.  Here we prove that result.

Recall that the matrix $\mathbf{V}_\mathrm{ov}$ contains the amplitudes defining $V_1$.  From Thouless' theorem,
\begin{equation}
\mathbf{V}_\mathrm{ov} = \mathbf{R}_\mathrm{oo}^{-1} \, \mathbf{R}_\mathrm{ov},
\end{equation}
where $\mathbf{R}_\mathrm{oo}$ and $\mathbf{R}_\mathrm{ov}$ are respectively the occupied-occupied and occupied-virtual blocks of $\mathbf{R}$ (and similarly, subscripts ``vv'' and ``vo'' would denote the virtual-virtual and virtual-occupied blocks).  This means that
\begin{subequations}
\begin{align}
\frac{\mathrm{d}\mathbf{V}_\mathrm{ov}}{\mathrm{d}\beta} 
 &= \frac{\mathrm{d} \mathbf{R}_\mathrm{oo}^{-1}}{\mathrm{d}\beta} \, \mathbf{R}_\mathrm{ov}
  + \mathbf{R}_\mathrm{oo}^{-1} \, \frac{\mathrm{d}\mathbf{R}_\mathrm{ov}}{\mathrm{d}\beta}
\\
 &= -\mathbf{R}_\mathrm{oo}^{-1} \, \frac{\mathrm{d}\mathbf{R}_\mathrm{oo}}{\mathrm{d}\beta} \, \mathbf{R}_\mathrm{oo}^{-1} \, \mathbf{R}_\mathrm{ov}
  + \mathbf{R}_\mathrm{oo}^{-1} \, \frac{\mathrm{d}\mathbf{R}_\mathrm{ov}}{\mathrm{d}\beta}
\\
 &= -\mathbf{R}_\mathrm{oo}^{-1} \, \frac{\mathrm{d}\mathbf{R}_\mathrm{oo}}{\mathrm{d}\beta} \, \mathbf{V}_\mathrm{ov}
  + \mathbf{R}_\mathrm{oo}^{-1} \, \frac{\mathrm{d}\mathbf{R}_\mathrm{ov}}{\mathrm{d}\beta}.
\end{align}
\label{Eqn:DerivV}
\end{subequations}

The rotation matrix $\mathbf{R}$ is
\begin{equation}
\mathbf{R} = \mathrm{e}^{-\mathrm{i} \, \beta \, \mathbf{S}_y}
\end{equation}
where recall that $\mathbf{S}_y$ is the matrix representation of the operator $S_y$.  Its derivative with respect to $\beta$ is therefore
\begin{equation}
\frac{\mathrm{d}\mathbf{R}}{\mathrm{d}\beta} = -\mathrm{i} \, \mathbf{R} \, \mathbf{S}_y \equiv \mathbf{R} \, \mathbf{A}
\end{equation}
where for convenience we have defined
\begin{equation}
\mathbf{A} = -\mathrm{i} \, \mathbf{S}_y.
\end{equation}
Inserting the derivative of $\mathbf{R}$ into Eqn. \ref{Eqn:DerivV} gives us
\begin{subequations}
\begin{align}
\frac{\mathrm{d}\mathbf{V}_\mathrm{ov}}{\mathrm{d}\beta} 
 &= -\mathbf{R}_\mathrm{oo}^{-1} \, \left(\mathbf{R}_\mathrm{oo} \, \mathbf{A}_\mathrm{oo} + \mathbf{R}_\mathrm{ov} \, \mathbf{A}_\mathrm{vo}\right) \, \mathbf{V}_\mathrm{ov}
\\
 &+ \mathbf{R}_\mathrm{oo}^{-1} \, \left(\mathbf{R}_\mathrm{oo} \, \mathbf{A}_\mathrm{ov} + \mathbf{R}_\mathrm{ov} \, \mathbf{A}_\mathrm{vv}\right)
\nonumber
\\
 &= -\mathbf{A}_\mathrm{oo} \, \mathbf{V}_\mathrm{ov} - \mathbf{V}_\mathrm{ov} \, \mathbf{A}_\mathrm{vo} \, \mathbf{V}_\mathrm{ov}
\\
 &+ \mathbf{A}_\mathrm{ov} + \mathbf{V}_\mathrm{ov} \, \mathbf{A}_\mathrm{vv},
\nonumber
\end{align}
\end{subequations}
which is the occupied-virtual block of the operator $\mathrm{e}^{V_1} \, A \, \mathrm{e}^{-V_1}$.  Accordingly,
\begin{equation}
\frac{\mathrm{d}\hfill}{\mathrm{d}\beta} V_1 = -\mathrm{i} \, \left(\mathrm{e}^{V_1} \, S_y \, \mathrm{e}^{-V_1}\right)_d
\end{equation}
where the subscript ``d'' means we keep the deexcitation part only and where we have used that $A = -\mathrm{i} \, S_y$.

\bibliography{PUCC}
\end{document}